\documentclass[twocolumn]{article}

\usepackage{lipsum}
\usepackage{amsmath}
\usepackage{amssymb}
\usepackage{graphicx}
\usepackage{hyperref}
\usepackage{geometry}
\usepackage{subcaption}
\usepackage{caption}
\usepackage{adjustbox}
\usepackage{float}
\usepackage{authblk}
\usepackage{rotating}
\usepackage{booktabs}
\usepackage{multirow}
\usepackage{tabularx}
\usepackage{longtable}
\usepackage{wrapfig}
\usepackage{xcolor}
\usepackage{soul}

\title{\textbf{A Walk across Europe: Development of a high-resolution walkability index}}

\author{Nishit Patel\textsuperscript{1,2,3}\thanks{Corresponding author, email: \texttt{n.t.patel@amsterdamumc.nl}},
       Hoang-Ha Nguyen\textsuperscript{1,2,3},
       Jet van de Geest\textsuperscript{1,2,3},
       Alfred Wagtendonk\textsuperscript{1,2,3},
       Mohan JS Raju\textsuperscript{1,2,3},
       Payam Dadvand\textsuperscript{4,5,6},
       Kees de Hoogh\textsuperscript{7,8},
       Marta Cirach\textsuperscript{4,5,6},
       Mark Nieuwenhuijsen\textsuperscript{4,5,6},
       Thao Minh Lam\textsuperscript{1,2,3},
       Jeroen Lakerveld\textsuperscript{1,2,3}
       }

\date{}

\begin{document}
\maketitle
\footnotetext[1]{Department of Epidemiology \& Data Science, Amsterdam UMC, location Vrije Universiteit, Amsterdam, the Netherlands}
\footnotetext[2]{Amsterdam Public Health research institute, Amsterdam UMC, Amsterdam, the Netherlands}
\footnotetext[3]{Upstream Team, \texttt{www.upstreamteam.nl}}
\footnotetext[4]{ISGlobal, Barcelona, Spain}
\footnotetext[5]{Universitat Pompeu Fabra (UPF), Barcelona, Spain}
\footnotetext[6]{CIBER Epidemiología y Salud Pública (CIBERESP)}
\footnotetext[7]{Swiss Tropical and Public Health Institute, Allschwil, Switzerland}
\footnotetext[8]{University of Basel, Basel, Switzerland}

\begin{abstract}
Physical inactivity significantly contributes to obesity and other non-communicable diseases, yet efforts to increase population-wide physical activity levels have met with limited success. The built environment plays a pivotal role in encouraging active behaviors like walking. Walkability indices, which aggregate various environmental features, provide a valuable tool for promoting healthy, walkable environments. However, a standardized, high-resolution walkability index for Europe has been lacking. This study addresses that gap by developing a standardized, high-resolution walkability index for the entire European region. Seven core components were selected to define walkability: walkable street length, intersection density, green spaces, slope, public transport access, land use mix, and 15-minute walking isochrones. These were derived from harmonized, high-resolution datasets such as Sentinel-2, NASA’s elevation models, OpenStreetMap, and CORINE Land Cover. A 100 m × 100 m hierarchical grid system and advanced geospatial methods, like network buffers and distance decay, were used at scale to efficiently model real-world density and proximity effects. The resulting index was weighted by population and analyzed at different spatial levels using visual mapping, spatial clustering, and correlation analysis. Findings revealed a distinct urban-to-rural gradient, with high walkability scores concentrated in compact urban centers rich in street connectivity and land use diversity. The index highlighted cities like Barcelona, Berlin, Munich, Paris, and Warsaw as walkability leaders. This standardized, high-resolution walkability index serves as a practical tool for researchers, planners, and policymakers aiming to support active living and public health across diverse European contexts.
\end{abstract}

\begin{figure*}[t]
    \centering 
    \includegraphics[width=\textwidth]{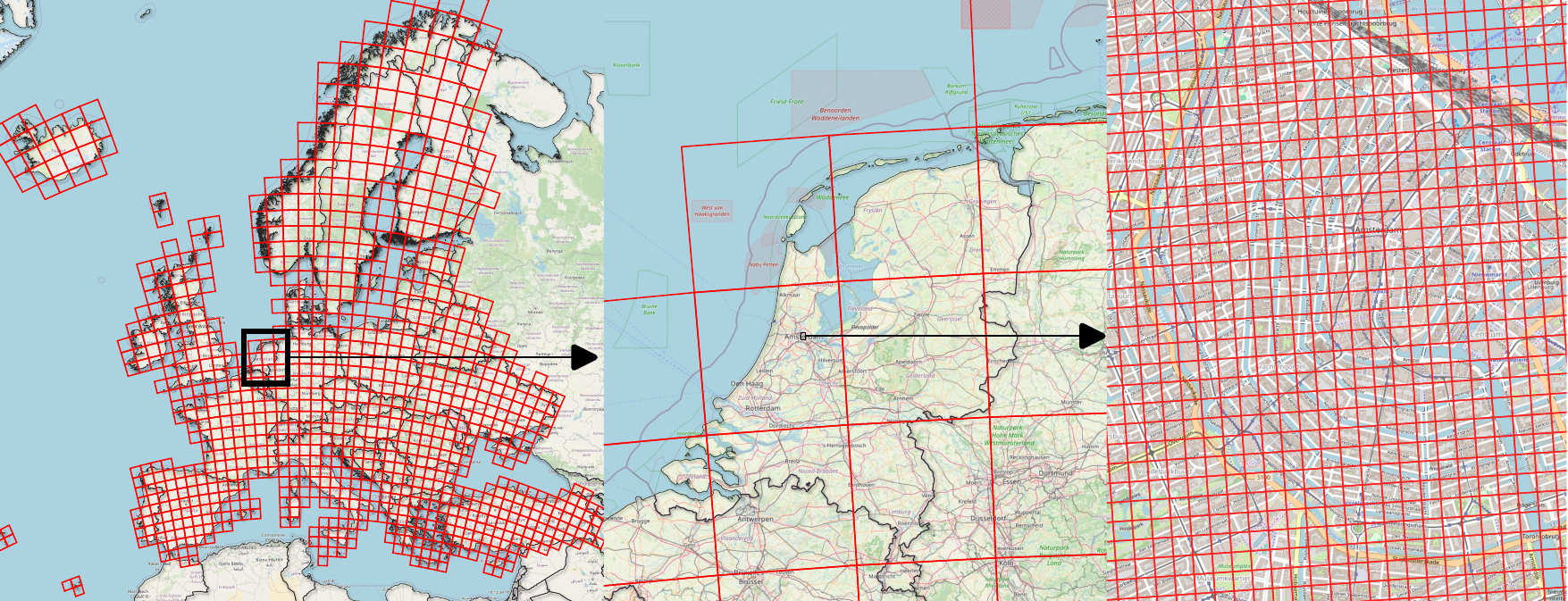}
    \caption{Hierarchical grid system. \textbf{Left panel}: Europe divided into 100 km x 100 km grids, \textbf{Middle panel}: Zoomed-in version of left panel over the Netherlands, \textbf{Right panel}: Zoomed-in version of middle panel over Amsterdam with 100 m x 100 m grids (neighborhoods)}
    \label{fig:1}
\end{figure*}

\section{Introduction}
Engaging in regular physical activity significantly lowers the risk of non-communicable diseases such as obesity, type 2 diabetes and cardiovascular diseases \cite{cleven2020association, lee2012effect, who2020guidelines}. However, over the past decades, levels of physical activity have been on a steady decline globally \cite{who2020guidelines}. Efforts to promote physical activity behaviors through individual-level interventions have shown limited success in achieving population-wide improvements \cite{strain2024national}. In response to this, research has increasingly focused on upstream factors that influence physical activity behaviors on population level. The context in which physical activity takes place, such as the built environment, has been reported to play a role in shaping individuals’ routes, routines, and physical activity habits \cite{lee2008neighbourhood, karmeniemi2018built}. For example, features such as street connectivity, mixed land use, and proximity to amenities have shown to impact the likelihood of engaging in active behaviors like walking \cite{mccormack2011search, wendel2007potential}.

Many studies, however, considered single environmental factors in relation to walking behavior, which oversimplifies the complexities of context in which individuals live and engage in physical activity, as they are simultaneously influenced by multiple environmental characteristics. To address this, walkability indices have emerged as robust combined measures that integrate key built environment features influencing walking and other types of physical activity. Commonly included environmental components encompass residential density, land use mix, street connectivity, the presence of green spaces, terrain slope, safety considerations, and proximity to key amenities such as schools and community centers. These factors are typically combined using arithmetic formulas to derive a composite walkability score.

Walkability indices have been widely utilized in research and have been associated with higher levels of walking activity \cite{WATSON2020106122}, overall physical activity \cite{rundle2019development, wei2016walkability, frank2005linking}, and various downstream health outcomes \cite{sundquist2015neighborhood}. Beyond their public health applications, walkability metrics have gained traction in urban and regional planning. Recent frameworks such as the “15-minute city” have positioned walkability as a central pillar in climate adaptation and sustainable urban development, emphasizing proximity to daily needs as a driver of low-carbon, health-promoting lifestyles \cite{bruno2024universal, fina2022walk}. In line with this shift, urban planners increasingly adopt a “pedestrian-first” approach to street design and land use regulation \cite{huang2024comprehensive}. These policy-driven applications highlight the importance of walkability indices not only for individual behavior analysis but also for benchmarking progress toward compact, inclusive, and climate-resilient cities.
However, the operationalization of walkability indices often lacks a standardized methodology \cite{venerandi2024walkability, shashank2019unpacking}. This variability is not unexpected, as most existing studies have constructed indices at local scales, heavily relying on region-specific datasets and local knowledge. Consequently, these locally developed indices are often unsuitable for broader regional or continental applications. Expanding walkability assessments to a larger spatial scale, such as the entire European region, facilitates standardized evaluations across diverse areas, providing policymakers with actionable benchmarks and researchers with opportunities for large-scale and comparative studies.

To this end, developing a Europe-wide, high-resolution walkability index addresses a critical gap in harmonized spatial indicators needed to assess urban accessibility, environmental equity, and active mobility infrastructure across diverse regions. Such indicators are increasingly essential for tracking progress toward EU-wide goals related to climate resilience, regional cohesion, and healthy urban living \cite{eu2021urbanmobility}. Our approach addresses this need by providing a reproducible methodology that supports consistent spatial comparisons across countries and administrative levels, while remaining grounded in open data and scalable tools.

Nonetheless, developing such indices poses significant methodological challenges, particularly in balancing the need for large-scale comparability with sufficient local granularity. Recent advances in computational capacity and the availability of standardized, high-resolution datasets — such as satellite imagery, OpenStreetMap (OSM), and other spatial resources — now enable the mapping of walkability across extensive areas without sacrificing detail. In this context, the present study aimed to develop a standardized, high-resolution walkability index (100 m x 100 m) for the whole of Europe by leveraging these datasets and advanced geospatial analysis techniques. Additionally, to extract meaningful insights, we analyzed the developed index at multiple spatial resolutions using population-weighted aggregation methods to explore spatial patterns relevant to governance and urban planning. To further support policymakers and researchers, we also provide open access to the resulting maps (through an interactive web atlas) and the underlying computational framework at \url{https://github.com/ohheynish/walkability-obct}.

\section{Methods}

The walkability index presented here was developed in the context of the European Commission-funded project OBCT (Obesity: Biological, socioCultural, and environmental risk Trajectories) \cite{lam2025obct}, and builds further on a previously proposed European-wide walkability measure developed in the EXPANSE project \cite{de2025europe, vlaanderen2021developing}.

\subsection{Study Area}\label{sec:study_area}
To develop a standardized walkability index for the European region (EU-27 plus Norway, United Kingdom, Switzerland, Iceland, Liechtenstein, Turkey, Albania, Montenegro, North Macedonia, Serbia, and Kosovo), we implemented a hierarchical grid system, as shown in Figure \ref{fig:1}. First, we divided the European region into large 100 km × 100 km grids (left-most panel in Figure \ref{fig:1}), which were then further subdivided into smaller 100 m × 100 m grids (right-most panel in Figure \ref{fig:1}), referred to as neighborhoods. This structured methodology was chosen to ensure consistency across the study area while optimizing computational efficiency (for instance, regular gridded vector shapes can be converted into arrays, enabling faster processing). This approach resulted in approximately 1 billion neighborhoods across the entire European region, spanning around 1,000 large-scale 100 km × 100 km grids. To generate these grids and operationalize all walkability components within our geoprocessing framework, we utilized the ETRS89-LAEA map projection (EPSG:3035), ensuring compatibility with standardized EU-wide data sources, such as Eurostat \footnote{\url{https://ec.europa.eu/eurostat/web/gisco/overview}}. 

\subsection{Walkability Components}
We selected a broad range of components to comprehensively characterise walkability, based on previously developed walkability indices \cite{lam2022development} as well as on the availability of Europe-wide standardisable data. To this end, the data on following seven components were collected for each 100 m x 100 m neighborhood: the length of walkable streets, the number of street intersections, the presence of green spaces, topographical variations, the number of public transport options, diversity in land use, and the area of 15 minute walking isochrones. These components are further detailed below:

\subsubsection{Street Walk Length (\texttt{SWL})}
\texttt{SWL} quantifies the total length of pedestrian streets within a 100 m × 100 m neighborhood. Initially, we modeled a comprehensive pedestrian street network for each 100 km × 100 km grid using the OSMnx\footnote{\url{https://github.com/gboeing/osmnx}} python library. Subsequently, using the GeoPandas\footnote{\url{https://github.com/geopandas/geopandas}} library, we calculated the total length of the walkable streets within each 100 m × 100 m neighborhood polygon.

\subsubsection{Street Intersections (\texttt{SI})}
To calculate the number of walkable street intersections within each neighborhood, we followed a similar methodology to the Street Walk Length component detailed above, starting with the construction of a walking network for the 100 km × 100 km grid via the OSMnx library. After establishing the network, we counted nodes where three or more streets intersect (three or more streets outgoing or incoming) within each 100 m × 100 m grid, providing a measure of intersection density. 


\subsubsection{Green Spaces (\texttt{NDVI} \& \texttt{GS})}
\label{sec:green_spaces}

The extent of green spaces within each neighborhood was assessed using a hybrid approach that combines satellite-derived vegetation indices with geographic information from OSM to enhance semantic precision and walkability relevance.

First, we calculated the Normalized Difference Vegetation Index (NDVI) using Sentinel-2 imagery, which offers 10 m spatial resolution. NDVI values for each 100 m × 100 m neighborhood were computed by averaging all underlying 10 m pixels falling within each grid cell. To ensure consistent and high-quality input, we selected imagery from the spring and summer seasons of 2018 and applied preprocessing steps such as cloud removal and temporal compositing. These operations were conducted using the Python API of Google Earth Engine\footnote{\url{https://github.com/google/earthengine-api}}.

To complement the NDVI-based assessment and differentiate between general vegetation and walkable green spaces (such as parks and recreational grounds), we also extracted green space polygons from OSM. Specifically, we filtered for publicly accessible green land use categories that are likely to represent functional, walkable areas. Private or inaccessible areas were excluded to ensure that only publicly usable and potentially walkable spaces were included.

\subsubsection{Slope (\texttt{SLOPE})}
The slope within each neighborhood was calculated using NASA's Digital Elevation Model (DEM), which offers a 30 m spatial resolution. \texttt{SLOPE} metric was derived by averaging the elevation change across the pixels within each 100 m × 100 m area. This data was accessed and processed through the Google Earth Engine's Python API, enabling consistent and reliable slope calculations across different neighborhoods. 

\subsubsection{Public Transport Stops (\texttt{PT})}
We assessed public transport accessibility by quantifying the number of transit options reachable on foot within each 100 m × 100 m neighborhood. This was calculated by counting the number of public transport stops, such as bus, tram, metro, subway, and train stations, located within each grid cell. Data for these stops were sourced from OSM using the Overpass API, with extraction facilitated through the Overpass\footnote{\url{https://github.com/mvexel/overpass-api-python-wrapper}} Python library. 

\subsubsection{Land Use Mix (\texttt{LUM})}
A Land use mix metric was used to evaluate the diversity of land uses within a neighborhood and was quantified using an entropy metric derived from the Copernicus CORINE land cover dataset\footnote{\url{https://developers.google.com/earth-engine/datasets/catalog/COPERNICUS_CORINE_V20_100m}}, which offers a 100 m spatial resolution. Broad land cover categories from the CORINE land cover dataset were first remapped into five primary classes that are frequently referenced in walkability research. This classification could help with assessing the diversity and functionality of land use, which is a key determinant of walkability. The categorization was as follows: 

\begin{itemize}
    \item Urban Fabric (categories 111, 112) was classified as \texttt{Class 1}, representing densely built residential or commercial areas.
    \item Industrial, Commercial, and Transport Units (categories 121-124) were grouped into \texttt{Class 2}, indicating regions predominantly used for industrial, commercial, or transportation purposes.
    \item Green Urban Areas (category 141) were designated as \texttt{Class 3}, highlighting green spaces within urban settings (e.g., urban parks).
    \item Sports and Leisure Facilities (category 142) fell under \texttt{Class 4}, reflecting areas dedicated to sports and recreational activities.
    \item Combined Agricultural and Natural Areas (categories 211-244, 311-324, 333, 511-512) were consolidated into \texttt{Class 5}, encompassing various forms of natural and agricultural land uses, from arable land and pastures to forests and water bodies.
\end{itemize}

After reclassification, entropy scores were calculated using the entropy function in Google Earth Engine. The resulting entropy scores were then mapped onto the corresponding 100 m x 100 m neighborhoods, effectively representing the (\texttt{LUM}) within those areas.

\subsubsection{15-minute walking isochrones (\texttt{ISO})}
This metric quantifies the amount of area that can be accessed by walking 15 minutes from a the centroid of a 100 m x 100 m neighborhood. The underlying rationale is that a larger walkable catchment area signifies greater accessibility to amenities, more opportunities for social and economic interaction, and better pedestrian infrastructure. The choice of a 15-minute walking threshold is grounded in the increasingly influential concept of the “15-minute city,” which promotes the idea that people should be able to meet most of their daily needs within a short walk from their homes. As discussed by \cite{bruno2024universal}, this paradigm not only supports sustainability and health through active mobility but also enhances quality of life by fostering local living and urban resilience. The use of this temporal threshold is therefore both a practical and conceptual alignment with emerging urban planning goals focused on compact, walkable, and livable neighborhood. Valhalla\footnote{\url{https://github.com/valhalla/valhalla}} routing engine was used to calculate these isochrones based on OSM's street network data. The walking speed parameter was kept at its default value of 5.1 km/h, meaning that a 15-minute walking time corresponded to an on-network distance of approximately 1,275 m.

\subsection{Walkability Index}
Since the characteristics of surrounding neighborhoods influence a target neighborhood’s walkability, developing a walkability index requires incorporating buffer areas of varying sizes to capture this effect. While Euclidean buffers are commonly used due to their computational simplicity, research suggests that street network buffers, which reflect actual walking paths, provide a more accurate representation of residential activity space \cite{li2022comparing, frank2021comparing, james2014effects, oliver2007comparing}. Therefore, for each 100 m × 100 m neighborhood, our index considered the distribution of walkability components within a 15-minute walking street network isochrone. This differs from the ‘Isochrones’ component, which only calculates the area accessible within a 15-minute walk; here, we analyze how various walkability components are distributed within that area. 

We also integrated distance decay functions into the isochrone buffers, assigning weights to environmental features based on their proximity to the target neighborhood. This reflects the principle that closer features have a stronger impact on our phenomenon of interest (i.e., walkability) than those farther away \cite{tobler1970computer}. To distribute these weights within the isochrone buffer, we apply a Gaussian function, which captures the varying spatial influence across the area. This method assigns higher weights to cells closer to the neighborhood of interest while gradually decreasing weights for those farther away. This weighted reduction process for the street network buffer can be formally expressed as:

\begin{equation}
\bar{c}_i(n_i) = \frac{\sum_{k \in K(n_i)} w_k \cdot c_i(k)}{\sum_{k \in K(n_i)} w_k}, \quad \forall c_i \in C, \, \forall n_i \in N
\label{eq:1}
\end{equation}
where:
\begin{itemize}
    \item \( \bar{c}_i(n_i) \): the weighted average of component \(c_i\) for neighborhood \(n_i\),
    \item \(c_i(k)\): the value of component \(c_i\) at grid cell \(k\),
    \item \(w_k\): the weight assigned to grid cell \(k\), determined by its distance from the center of the buffer,
    \item \(K(n_i)\): the set of grid cells within the 15-minute isochrone for neighborhood \(n_i\),
    \item \(C\): the set of walkability components,
    \item \(N\): the set of all neighborhoods.
\end{itemize}
 
After re-weighing the component values for all neighborhoods as indicated in Equation \eqref{eq:1}, these values were then standardized using z-score normalization:

\begin{equation}
z_{c_i}(n_i) = \frac{\bar{c}_i(n_i) - \mu_{c_i}}{\sigma_{c_i}}, \quad \forall c_i \in C, \, \forall n_i \in N
\label{eq:2}
\end{equation}
where:
\begin{itemize}
    \item \(z_{c_i}(n_i)\): the standardized value of component \(c_i\) for neighborhood \(n_i\),
    \item \(\mu_{c_i}\): the mean of component \(c_i\) across all neighborhoods,
    \item \(\sigma_{c_i}\): the standard deviation of component \(c_i\) across all neighborhoods.
\end{itemize}

Before compiling the standardized component scores into the final index, we assessed the correlation structure among the individual components at the \texttt{LAU} level to identify potential redundancy. This analysis revealed a very high correlation between street walkable length and intersection density (r = 0.95) (Figure \ref{fig:corr_matrix}), indicating that they capture largely overlapping aspects of street network design. To avoid over-representing this single environmental characteristic, we combined the two components using equal weights (0.5 each) in the final index. This adjustment ensures that highly similar features do not disproportionately influence the index and supports diversity in the types of environmental information captured.

Similarly, as described in Section \ref{sec:green_spaces}, we incorporated both \texttt{GS} and \texttt{NDVI} as complementary indicators of green environments. While \texttt{GS} reflects the availability of designated green areas, \texttt{NDVI} provides a continuous measure of vegetative presence, including informal greenery. To balance their contributions, these components were also averaged using equal weights (0.5 each), ensuring a more comprehensive representation of the greenness dimension.

Furthermore, the slope component ($z_{c_\texttt{SLOPE}}$) is incorporated with a negative weight to reflect its inverse relationship with walkability. While we acknowledge that uphill segments may be desirable in recreational or fitness contexts, our focus is to capture walkability from a general accessibility perspective. Steeper slopes are generally understood to reduce the comfort and practicality of walking, particularly for those with mobility constraints. Once the components were weighted, they were compiled into a comprehensive walkability index as:

\begin{align}
\text{Index}(n_i) &= \text{0.5}z_{c_\texttt{SWL}} + \text{0.5}z_{c_\texttt{SI}} + \text{0.5}z_{c_\texttt{GS}} + \text{0.5}z_{c_\texttt{NDVI}} \nonumber  \\ &\quad - z_{c_\texttt{SLOPE}} 
 + z_{c_\texttt{PT}} + z_{c_\texttt{LUM}} + z_{c_\texttt{ISO}}, \quad \forall n_i \in N
\label{eq:3}
\end{align}
where:
\begin{itemize}
    \item \(\text{Index}(n_i)\): walkability index for neighborhood \(n_i\).
\end{itemize}

To make the score more intuitive, we divide $\text{Index}(n_i)$ into deciles.

As discussed in Section \ref{sec:study_area}, the extensive coverage and relatively high granularity of our study area result in a substantial number of grids ($\sim$1 billion). To optimize processing efficiency and leverage the structured grid system, we implemented a Graphics Processing Unit (GPU)-based algorithm to construct our index after data collection.

\subsection{Statistical Analysis}
To derive meaningful insights from the constructed walkability index across Europe, we analyzed the index at four distinct spatial levels: \texttt{Countries} (n=38), \texttt{NUTS-3}\footnote{\url{https://ec.europa.eu/eurostat/web/gisco/geodata/statistical-units/territorial-units-statistics}} regions (n=1248), \texttt{LAU}\footnote{\url{https://ec.europa.eu/eurostat/web/gisco/geodata/statistical-units/local-administrative-units}} units (n=96927), and 100 m x 100 m neighborhoods for selected meteropolitan cities. This multi-scalar approach enables a comprehensive understanding of spatial patterns, variations, and trends in walkability across different levels of governance and urban planning. Analyzing walkability at these administrative and fine-grained levels supports cross-country comparisons, identifies regional disparities, and offers practical insights for urban policy, regional development strategies, and climate-resilient planning. Furthermore, analyzing walkability at 100m × 100m resolution enables detailed evaluations at the neighborhood level, supporting future linkage with individual-level health outcomes and the design of targeted public health interventions.

\subsubsection{Methodology for Spatial Aggregation}
For each administrative unit, we first select the constituent 100m x 100m grid neighborhoods that fall within the boundaries of polygons (regions) included in that level. To ensure comparability across different regions, we computed population-weighted versions of our walkability components based on the aggregated values obtained after Equation \eqref{eq:1}. This population-weighted transformation can be formally defined as:

\begin{equation}
\bar{c}_i^{\texttt{POP}} = \frac{\sum\bar{c}_i(n_i) \cdot p_i(n_i)}{\sum p_i(n_i)}, \quad \forall n_i \in N
\label{eq:4}
\end{equation}
where:
\begin{itemize}
    \item \( p_i(n_i) \): the population of the neighborhood \( n_i \),
    \item \( N \): the set of all neighborhoods within a polygon of the respective administrative unit.
\end{itemize}

The population data for each 100 m x 100 m neighborhood was obtained from the Global Human Settlement Layer (GHSL): Global Population Surfaces (2020) \cite{carioli2023ghs}. This dataset provides high-resolution population estimates, allowing for accurate weighing of walkability components.

\subsubsection{Stratification by Degree of Urbanization}
To further contextualize our findings, we classified each polygon within the administrative units based on its degree of urbanization, using the GHSL: Degree of Urbanization dataset \cite{carioli2023ghs}. There are 7 main classes included in this dataset: very low density rural, low density rural, rural cluster, suburban or peri-urban, semi-dense urban cluster, dense urban cluster, and urban center. This classification allowed us to segment regions into urban, suburban, and rural categories. We calculated correlation coefficients between individual walkability components and the overall walkability index for \texttt{LAU} administrative level. This correlation analysis provides insights into potential collinearity among components, which is crucial for assessing the redundancy of variables and optimizing the index structure.

\subsubsection{High-Resolution Mapping and Spatial Autocorrelation Analysis}
To illustrate the spatial distribution of walkability at the most granular level, we identified the 20 most populated \texttt{LAU} units (metropolitan cities) across Europe and mapped their walkability index at the original 100 m x 100 m resolution. Unlike the previously discussed population-weighted aggregation (Equation \eqref{eq:4}), these maps retain the original resolution to preserve local variations.

For these metropolitan cities, we also computed Moran's I (Equation \eqref{eq:5}). Moran's I quantifies the degree of spatial autocorrelation in the walkability index for each of the selected metropolitan cities. This global Moran’s I metric serves as an indicator of spatial clustering, revealing whether the index exhibits spatial dependency.

\begin{equation}
I = \frac{N}{W} \cdot \frac{\sum_{i} \sum_{j} w_{ij} (x_i - \bar{x}) (x_j - \bar{x})}{\sum_{i} (x_i - \bar{x})^2}
\label{eq:5}
\end{equation}
where:
\begin{itemize}
    \item \( N \): number of neighborhoods in the \texttt{LAU} unit (metropolitan city),
    \item \( W = \sum_{i} \sum_{j} w_{ij} \): sum of all spatial weights,
    \item \( x_i \): walkability value of neighborhood i,
    \item \( x_j \): walkability value of the neighboring neighborhood j,
    \item \( \bar{x} \): mean walkability of the \texttt{LAU} unit (metropolitan city),
    \item \( w_{ij} \): spatial weights between the neighborhoods i and j.
\end{itemize}

\begin{figure}[H]
    \centering 
    \includegraphics[width=0.90\columnwidth]{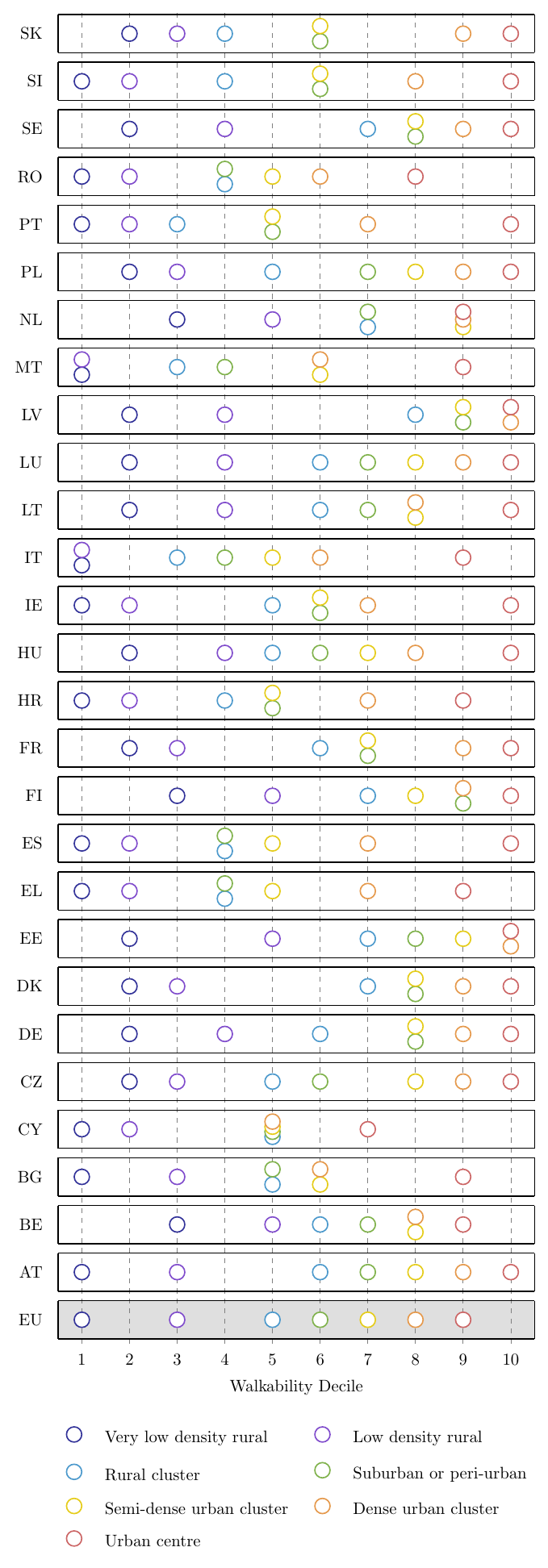}
    \caption{Comparison of walkability deciles across EU-27 countries, categorized by urbanization levels.
    Each horizontal bar represents a country and shows the distribution of urbanization clusters across walkability deciles. The EU baseline, outlined in gray, represents the population-weighted average of all 100 m × 100 m neighborhoods across Europe. This comparison highlights both within- and across-country disparities in walkability, stratified by urbanization level, revealing how urban structure and development patterns influence the accessibility of walkable environments.}
    \label{fig:strip_plot}
\end{figure}

Furthermore, to compare walkability distributions across urbanization gradients at this high-resolution, we generated population-weighted cumulative distribution plots of the walkability index stratified by degrees of urbanization. For each country, the walkability index was computed at 100 m x 100 m grid level using \ref{eq:3}. However, since grids span diverse urban-rural contexts at country scale, we developed per capita measures for components such as \texttt{SWL}, \texttt{SI}, \texttt{GS}, and \texttt{PT} prior to building the index. This per capita adjustment enables more accurate comparisons of walkability across diverse settlement types when analyzed at high spatial resolution and national scale.

\section{Results}
\begin{figure}[htbp]
    \centering 
    \includegraphics[width=0.90\columnwidth]{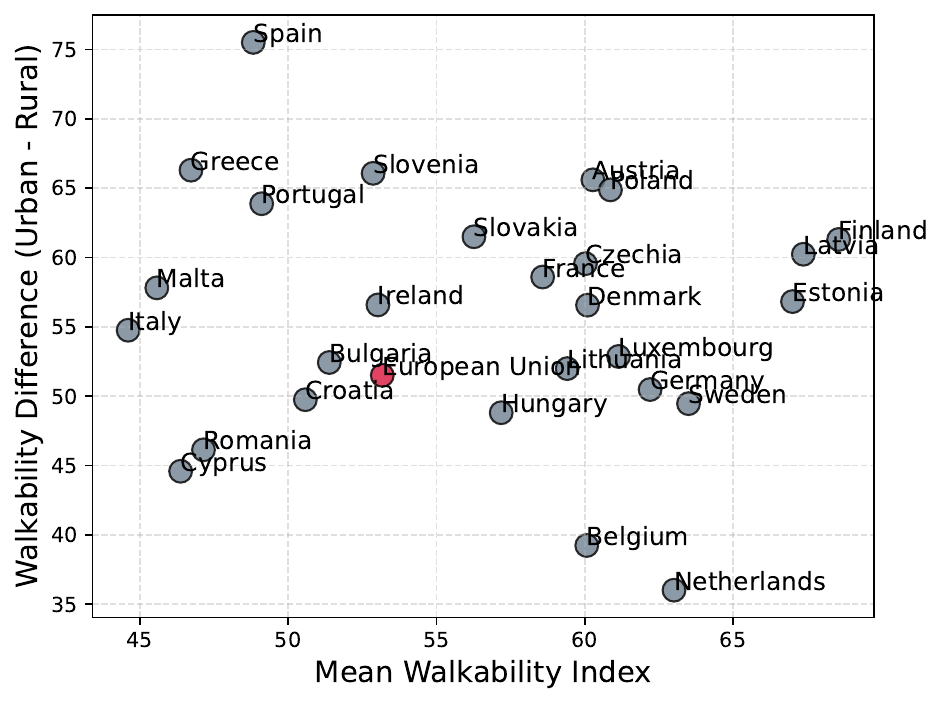}
    \caption{Relationship between the mean walkability index and the urban–rural walkability difference across European countries. The scatter plot illustrates how walkability varies between urban and rural areas within each country, highlighting disparities. For example, countries in the lower right corner (Belgium \& Netherlands) exhibit relatively higher overall walkability along with smaller urban-rural differences, suggesting a more equitable distribution of walkable environments.}
    \label{fig:disparity_plot}
\end{figure}

\begin{figure*}[t]
    \centering
    \includegraphics[width=\textwidth]{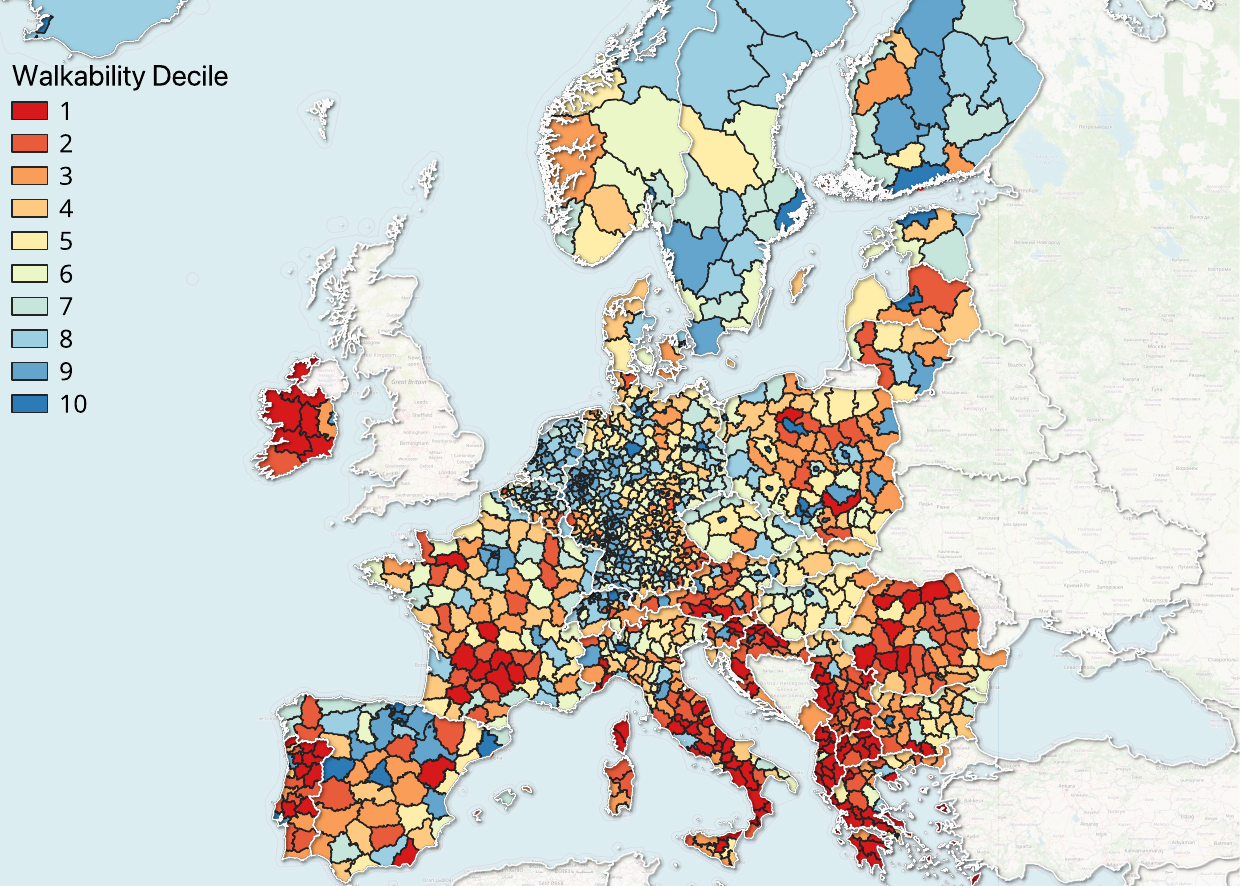}
    \caption{Walkability index map of \texttt{NUTS-3} units. Each unit reflects the population-weighted average walkability score of all 100 m × 100 m neighborhoods within that region. This level of aggregation provides a comparable overview across administrative regions. This map serves as a regional summary layer of the more detailed neighborhood-scale index developed in this study. An interactive version of this map with satellite and OSM base layers as well as additional interactive tools, is available \href{https://ohheynish.github.io/walkability_web_atlas}{here}.}
    \label{fig:nuts3_map}
\end{figure*}

Figure \ref{fig:strip_plot} depicts the average distribution of walkability across EU-27 countries, categorized by urbanization levels. The figure also includes the decile distribution of the walkability index for the entire EU, serving as a baseline for comparison with individual countries. For rural neighborhoods, several countries—Austria (AT), Belgium (BE), Germany (DE), Denmark (DK), Estonia (EE), Finland (FI), France (FR), Lithuania (LT), Luxemborg(LU), Latvia (LV), the Netherlands (NL), and Sweden (SE)—exceeded the EU baseline. Similarly, in dense urban clusters, Austria (AT), Czechia (CZ), Germany (DE), Denmark (DK), Estonia (EE), Finland (FI), France (FR), Luxembourg (LU), Latvia (LV), the Netherlands (NL), Poland (PL), Sweden (SE), and Slovakia (SK) ranked above the EU baseline. The walkability gradient followed a predictable urbanization pattern, with lower walkability in rural areas that progressively improved towards urban centers. This trend aligned with key contributing factors such as street walk length, 15-minute walking isochrones, street intersections, and public transport options, which consistently show higher values in urban environments.

Figure \ref{fig:disparity_plot} presents the relationship between the mean walkability index and the urban-rural walkbaility differnce across EU countries. The plot also reveals a consistent pattern in which urban areas exhibit higher walkability scores than rural areas in every country. However, the magnitude of this disparity varies considerably. Countries positioned toward the lower right corner of the plot, such as Belgium and the Netherlands, combine relatively high average walkability with smaller urban–rural differences. This suggests a more equitable distribution of walkability-enabling features across the urban–rural spectrum in these contexts. In contrast, countries toward the upper left corner, like Spain, tend to have both lower overall walkability and larger disparities, indicating that walkable environments are more concentrated in urban centers and less accessible in rural areas. These patterns reflect differences in urban planning, infrastructure investment, and land use integration across national contexts.

\begin{figure*}[t]
    \centering
    \begin{subfigure}[b]{0.32\textwidth}
        \includegraphics[width=\textwidth]{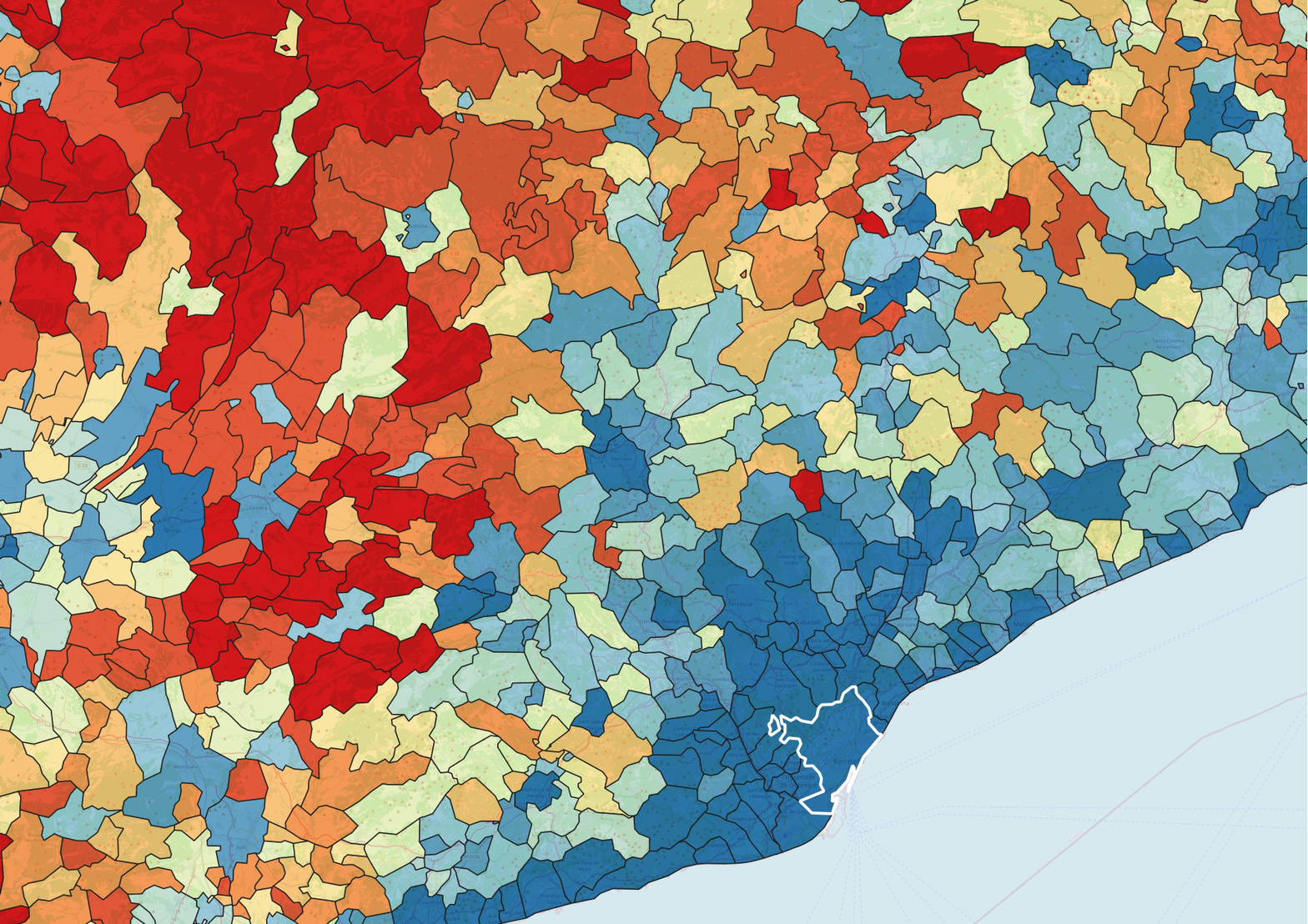}
        \caption{Barcelona}
    \end{subfigure}
    \begin{subfigure}[b]{0.32\textwidth}
        \includegraphics[width=\textwidth]{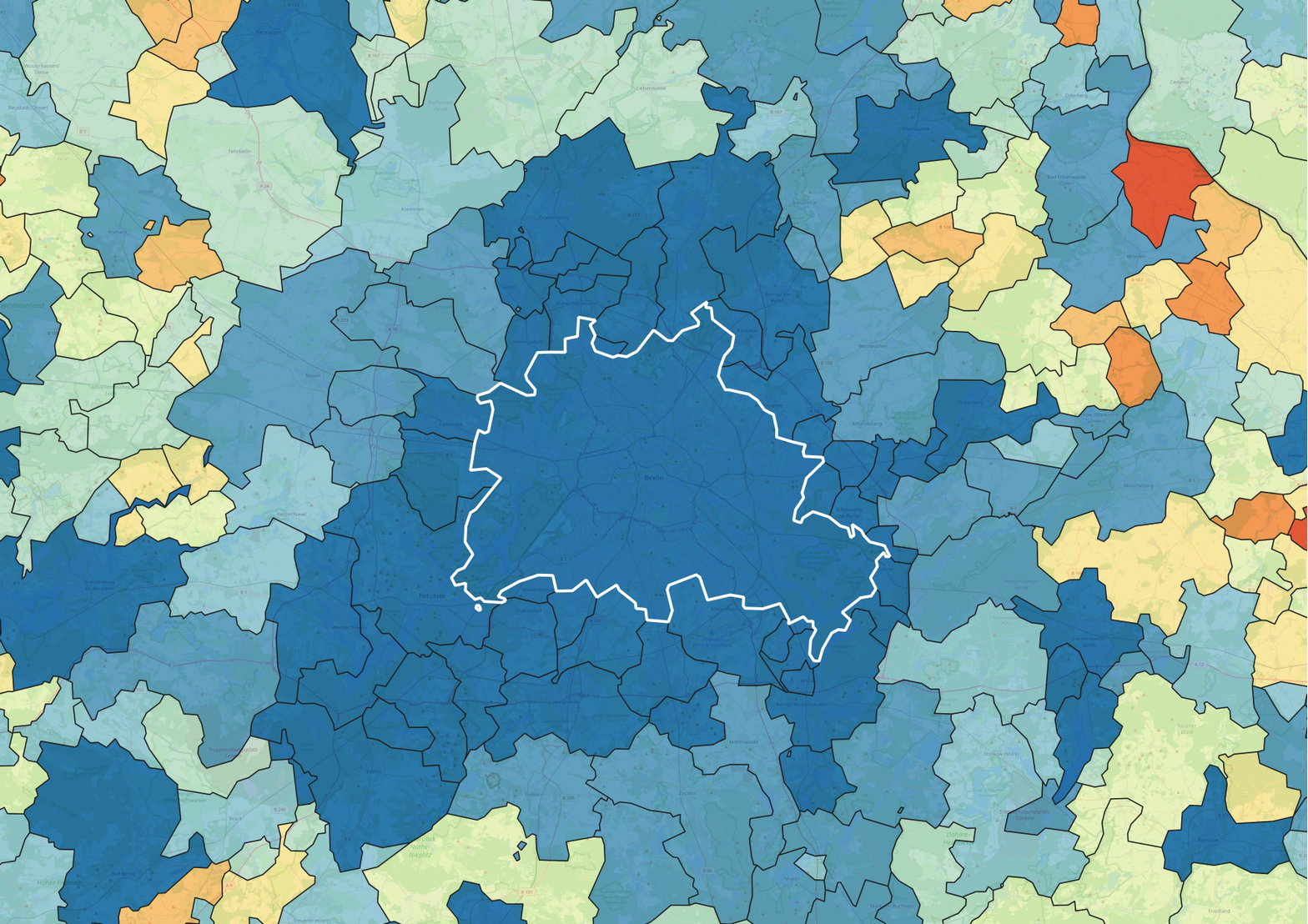}
        \caption{Berlin}
    \end{subfigure}
    \begin{subfigure}[b]{0.32\textwidth}
        \includegraphics[width=\textwidth]{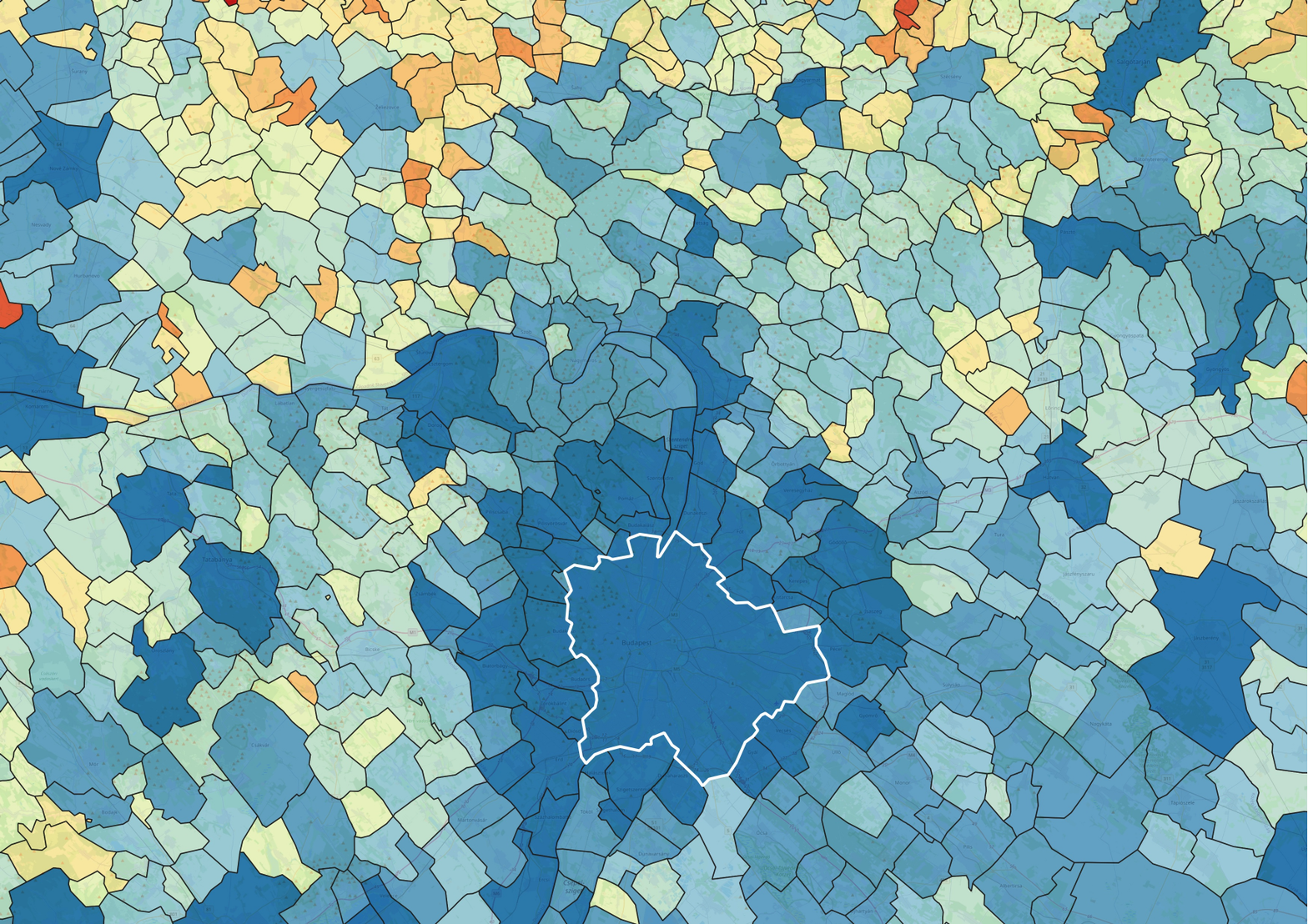}
        \caption{Budapest}
    \end{subfigure}
    
    \begin{subfigure}[b]{0.32\textwidth}
        \includegraphics[width=\textwidth]{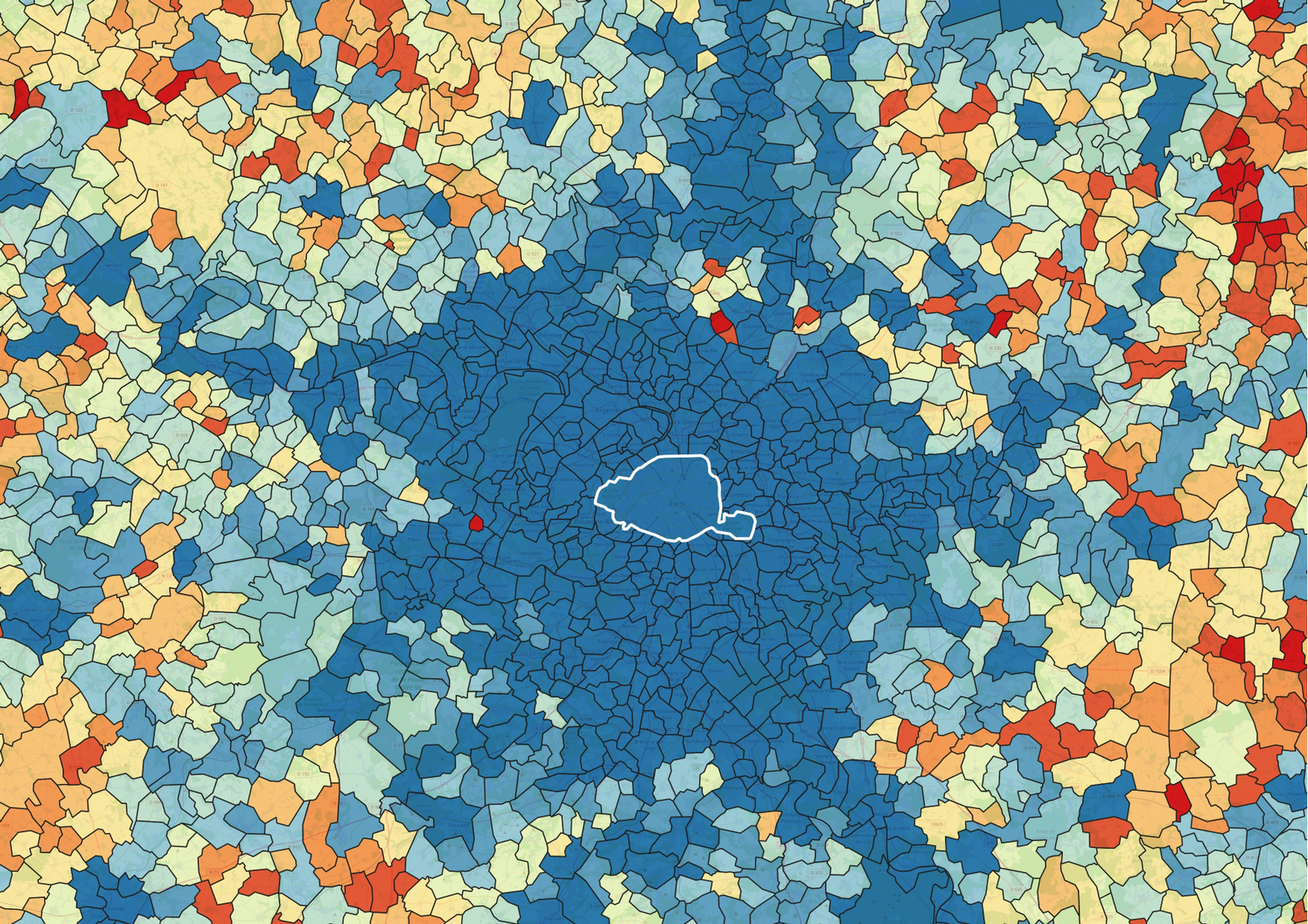}
        \caption{Paris}
    \end{subfigure}
    \begin{subfigure}[b]{0.32\textwidth}
        \includegraphics[width=\textwidth]{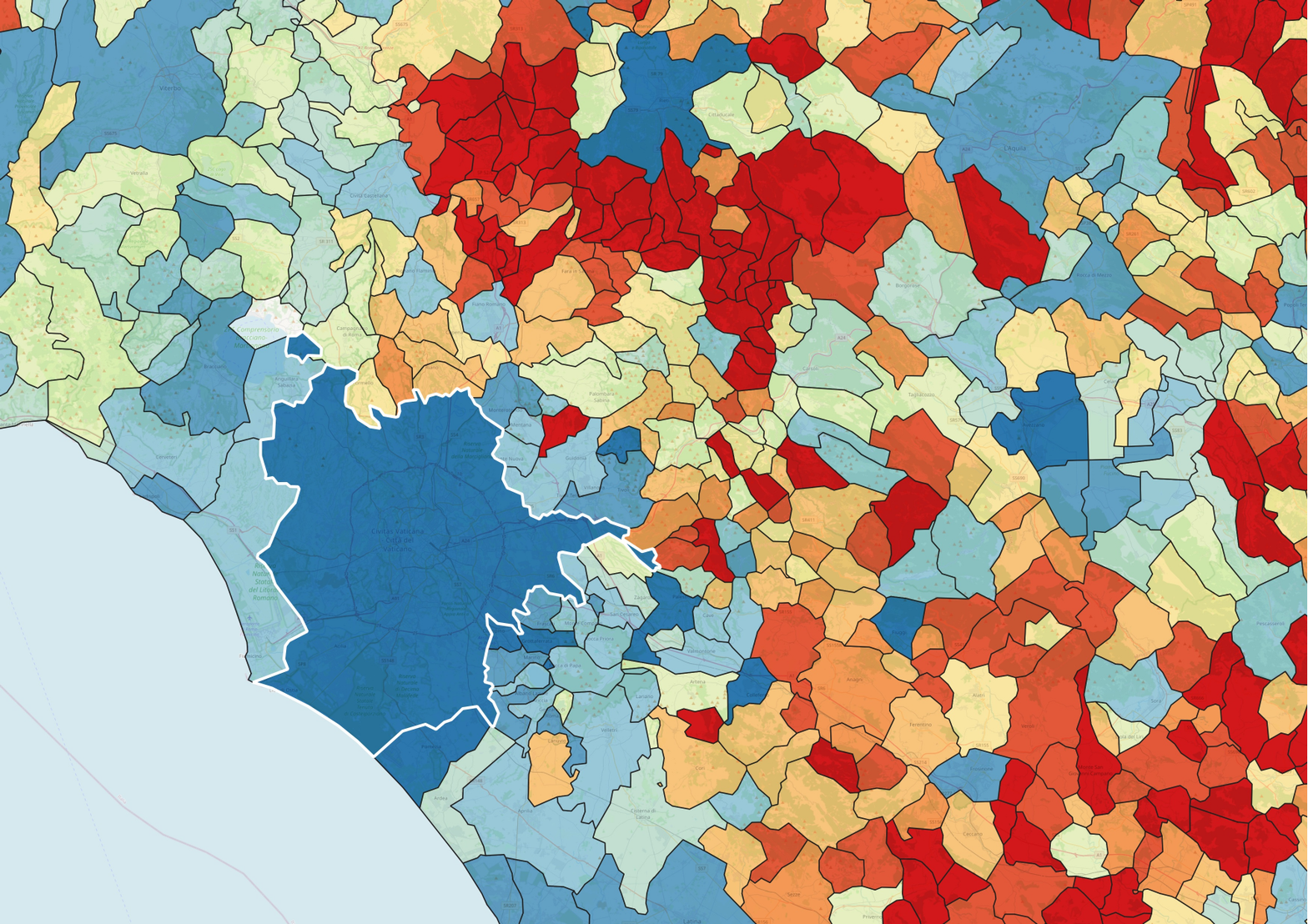}
        \caption{Rome}
    \end{subfigure}
    \begin{subfigure}[b]{0.32\textwidth}
        \includegraphics[width=\textwidth]{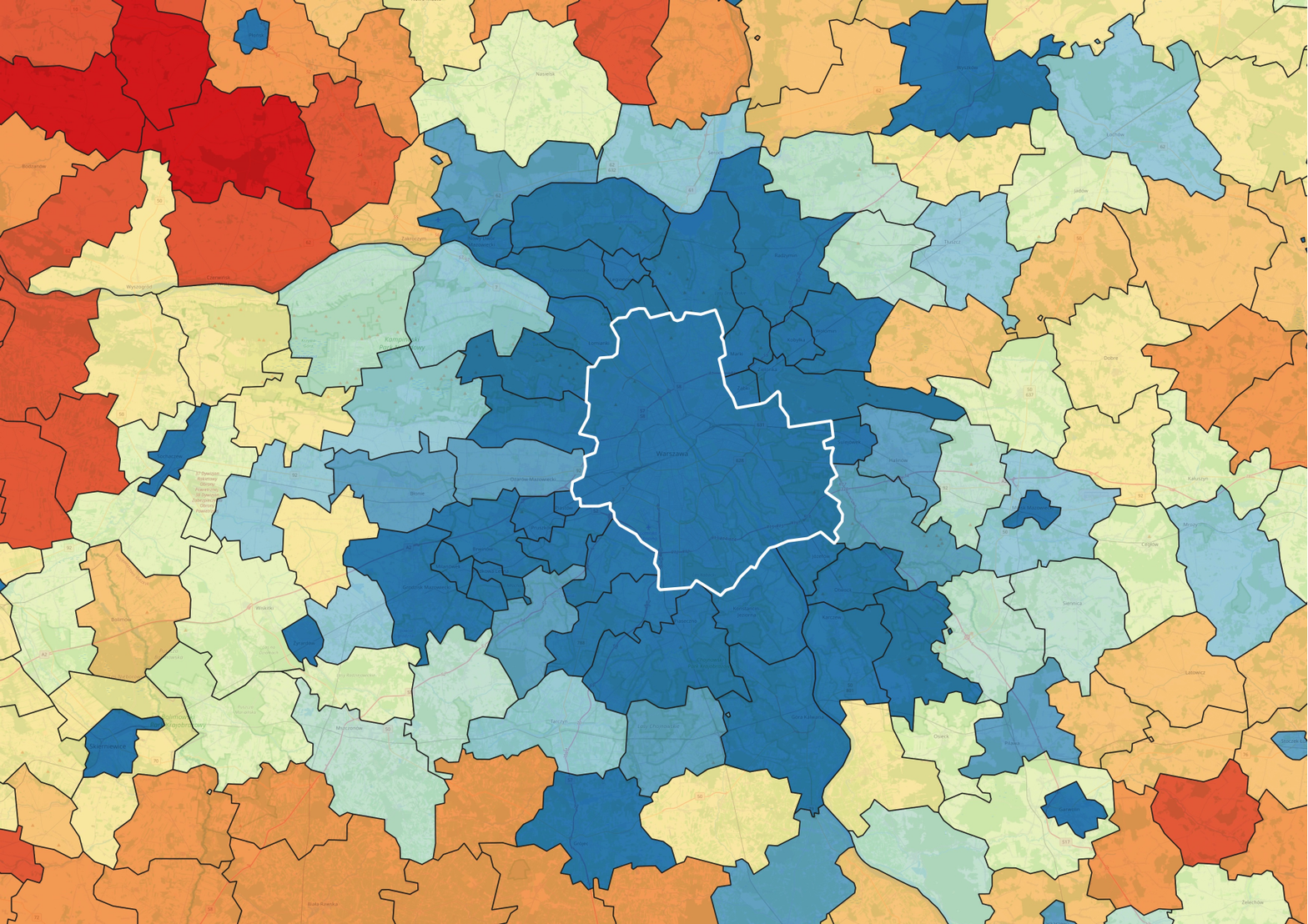}
        \caption{Warsaw}
    \end{subfigure}

    \begin{subfigure}[b]{\textwidth}
        \centering
        \includegraphics[width=0.7\textwidth]{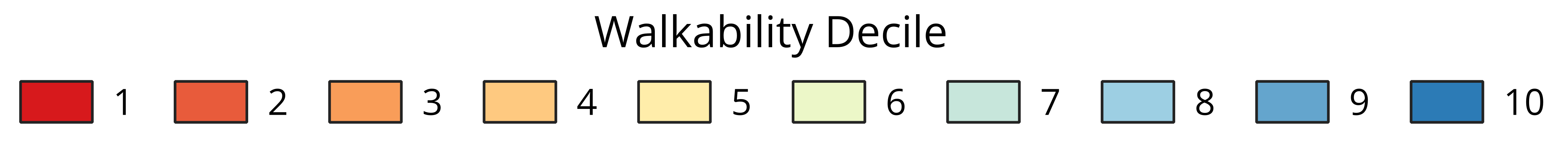}
    \end{subfigure}
    \caption{Zoomed-in walkability index maps of selected \texttt{LAU} units. Units outlined with white border represent cities mentioned in the sub-caption. Each unit reflects the population-weighted average walkability score of all 100 m × 100 m neighborhoods within that region. An interactive version of this map containing all \texttt{LAU} units with satellite and OSM base layers as well as additional interactive tools, is available \href{https://ohheynish.github.io/walkability_web_atlas}{here}.}
    \label{fig:lau_map}
\end{figure*}

A deeper exploration of the walkability patterns at the level of \texttt{NUTS-3} administrative units is presented in Figure \ref{fig:nuts3_map}, whereas, Figure \ref{fig:lau_map} shows the walkability patterns at \texttt{LAU} units for selected regions across Europe. A fully interactive version of these maps is available online \href{https://ohheynish.github.io/walkability_web_atlas}{here}.

\begin{figure}
    \centering
    \includegraphics[width=\linewidth]{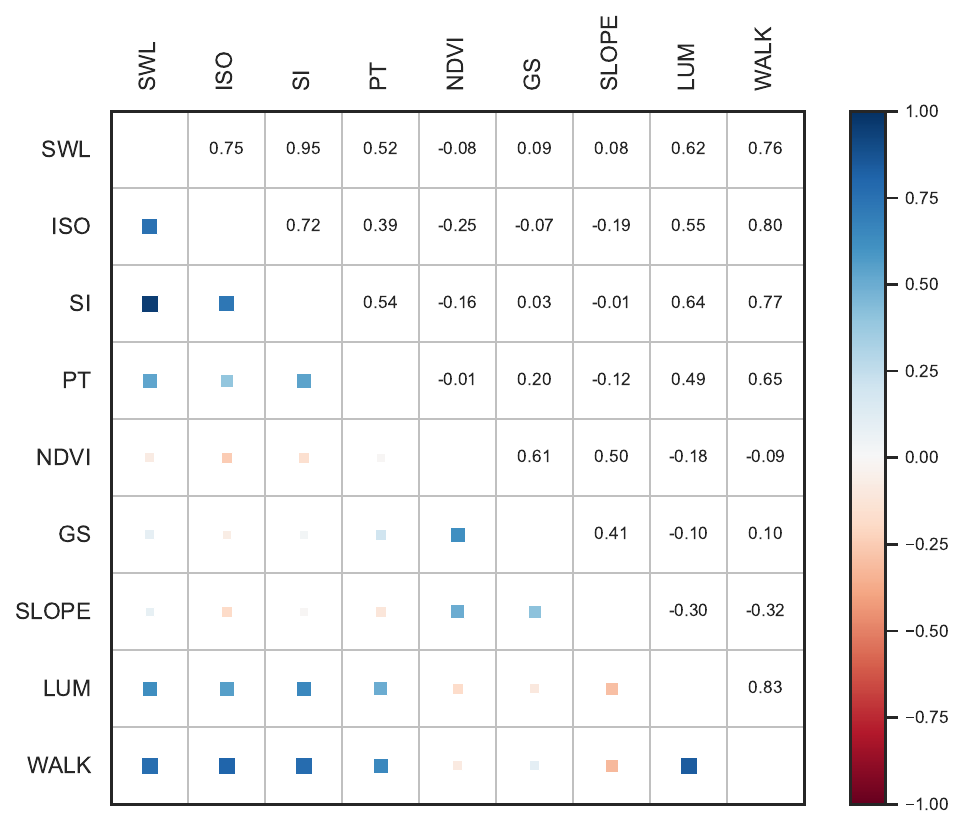}
    \caption{Correlation matrix calculated at \texttt{LAU} level. \texttt{SWL}: Street Walk Length, \texttt{SI}: Street Intersections, \texttt{GS}: Green Spaces, \texttt{SLOPE}: Slope, \texttt{PT}: Public Transport Stops, \texttt{LUM}: Land Use Mix, \texttt{ISO}: 15-min walking isochrones, \texttt{WALK}: Walkability Index.}
    \label{fig:corr_matrix}
\end{figure}

Figure \ref{fig:corr_matrix} presents the correlation matrix for walkability components at the \texttt{LAU} level. A strong correlation is observed between street intersections (\texttt{SI}) and street walk length (\texttt{SWL}), indicating that a higher intersection density is associated with an increased walkable street length. In contrast, green spaces (\texttt{GS}) shows a weak negative correlation with the other components, suggesting that as urban features intensify, the value of green spaces (\texttt{GS}) tends to decrease.

\begin{figure*}
    \centering 
    \includegraphics[width=\textwidth]{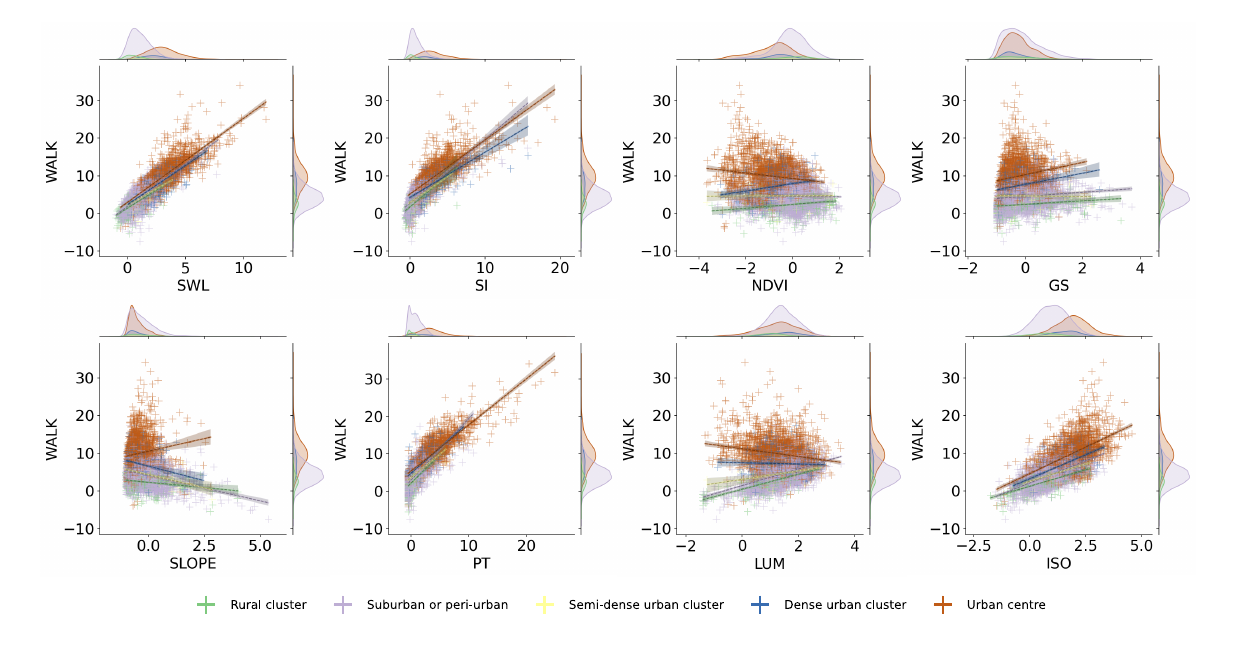}
    \caption{Jointplots with linear regression fit between individual components and the developed walkability index stratified by degree of urbanization. Unit of analysis: \texttt{LAU} units.}
    \label{fig:joint_plot}
\end{figure*}

Figure \ref{fig:joint_plot} displays joint plots of the individual components with walkability for \texttt{LAU} units, along with a fitted linear regression line. Interestingly, although \texttt{SLOPE} was modelled negatively for the overall walkability index (i.e., more elevation theoretically reduces walkability), the linear regression fit between slope (\texttt{SLOPE}) and the walkability index (\texttt{WALK}) for the sample of neighborhoods belonging to urban centers appears to be positive, as shown in Figure \ref{fig:joint_plot}. Similarly, the regression fit between green spaces (\texttt{GS}) and walkability index (\texttt{WALK}) in urban centers is negative, which seems counterintuitive. These exploratory findings provide insights into refining the index, potentially by employing different versions of Equation \eqref{eq:3} tailored to varying degrees of urbanization. 

\begin{figure*}
    \centering
    \begin{subfigure}[b]{0.45\textwidth}  
        \includegraphics[width=\textwidth]{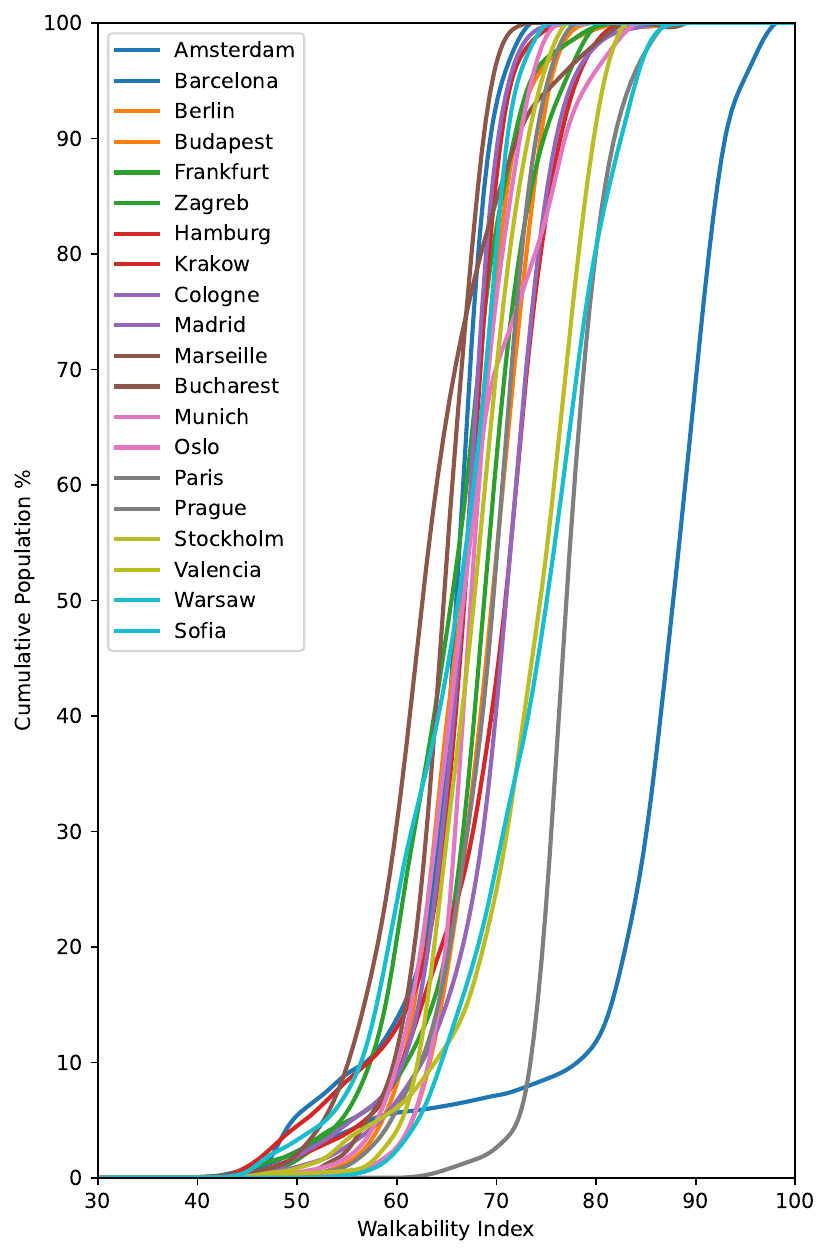}
        \caption{}
        \label{fig:curve}
    \end{subfigure}
    \begin{subfigure}[b]{0.45\textwidth}  
        \includegraphics[width=\textwidth]{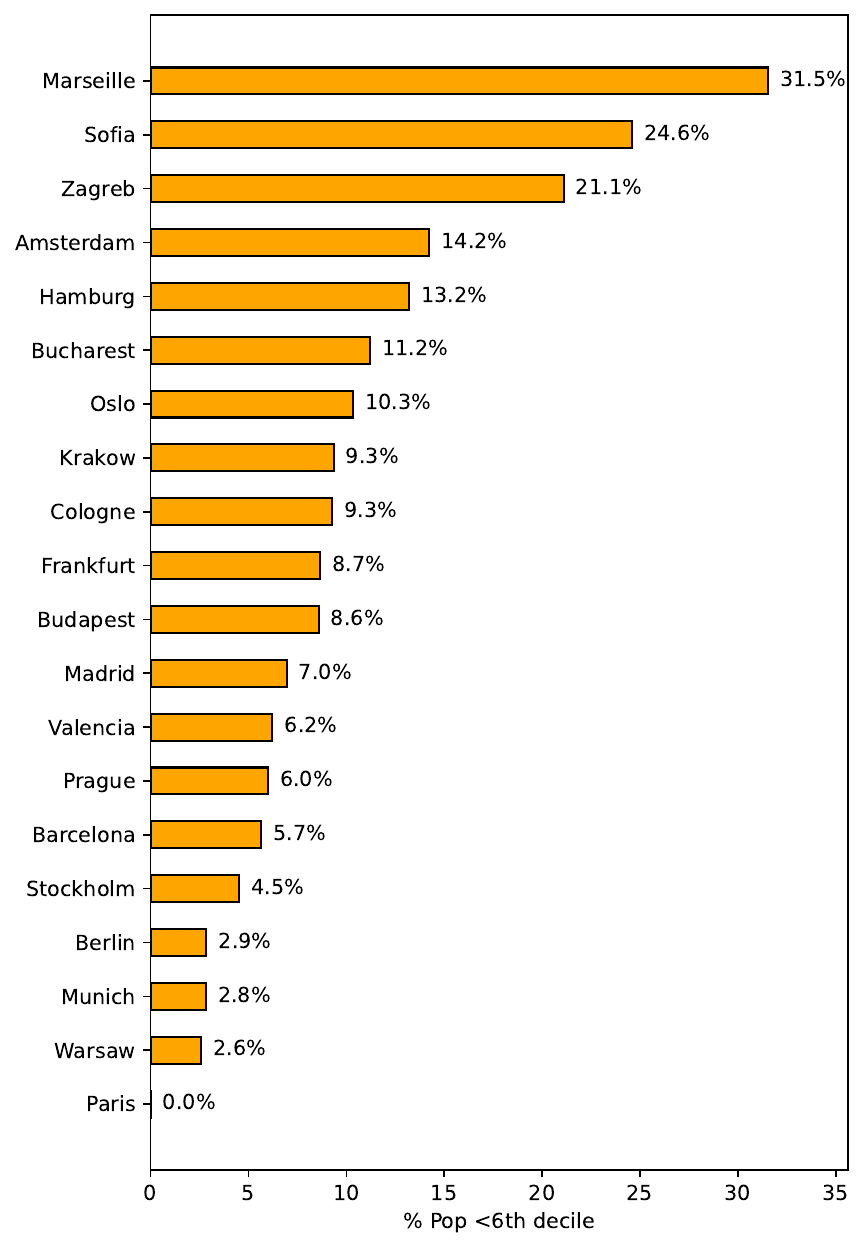}
        \caption{}
        \label{fig:barh}
    \end{subfigure}
    
    \caption{\textbf{(a)}: Percentage of people living below $6^\text{th}$ decile of walkability across various metropolitan cities across Europe, \textbf{(b)}: The cumulative distribution of the developed walkability index across population for top-20 most populated cities in Europe. Generally, the more right-shifted the curve, the higher the proportion of people living in higher walkability neighborhoods.}
    \label{fig:curve_barh}
\end{figure*}

\begin{figure*}[t]
    \centering
    \begin{subfigure}[b]{0.24\textwidth}
        \includegraphics[width=\textwidth]{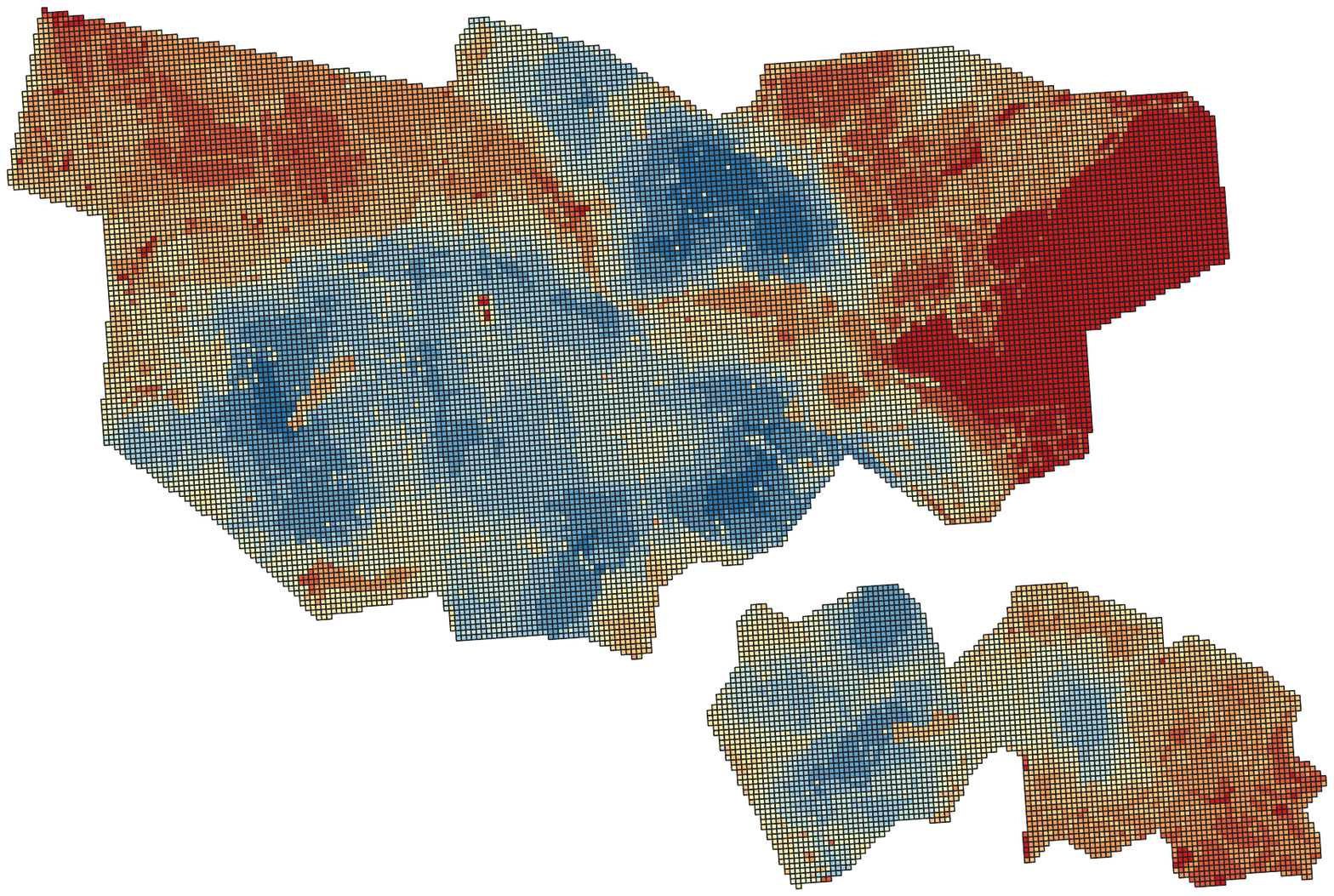}
        \caption{Amsterdam, $I=0.95$}
    \end{subfigure}
    \begin{subfigure}[b]{0.24\textwidth}
        \includegraphics[width=\textwidth]{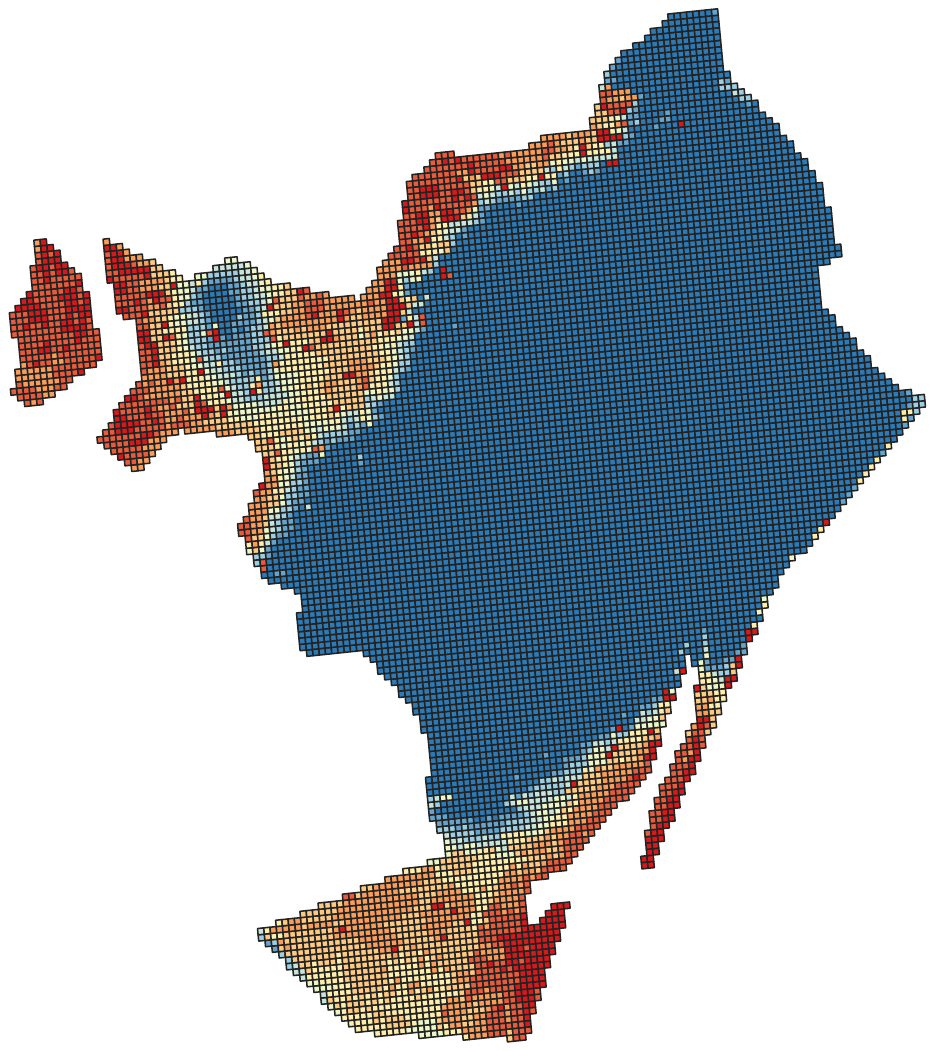}
        \caption{Barcelona, $I=0.97$}
    \end{subfigure}
    \begin{subfigure}[b]{0.24\textwidth}
        \includegraphics[width=\textwidth]{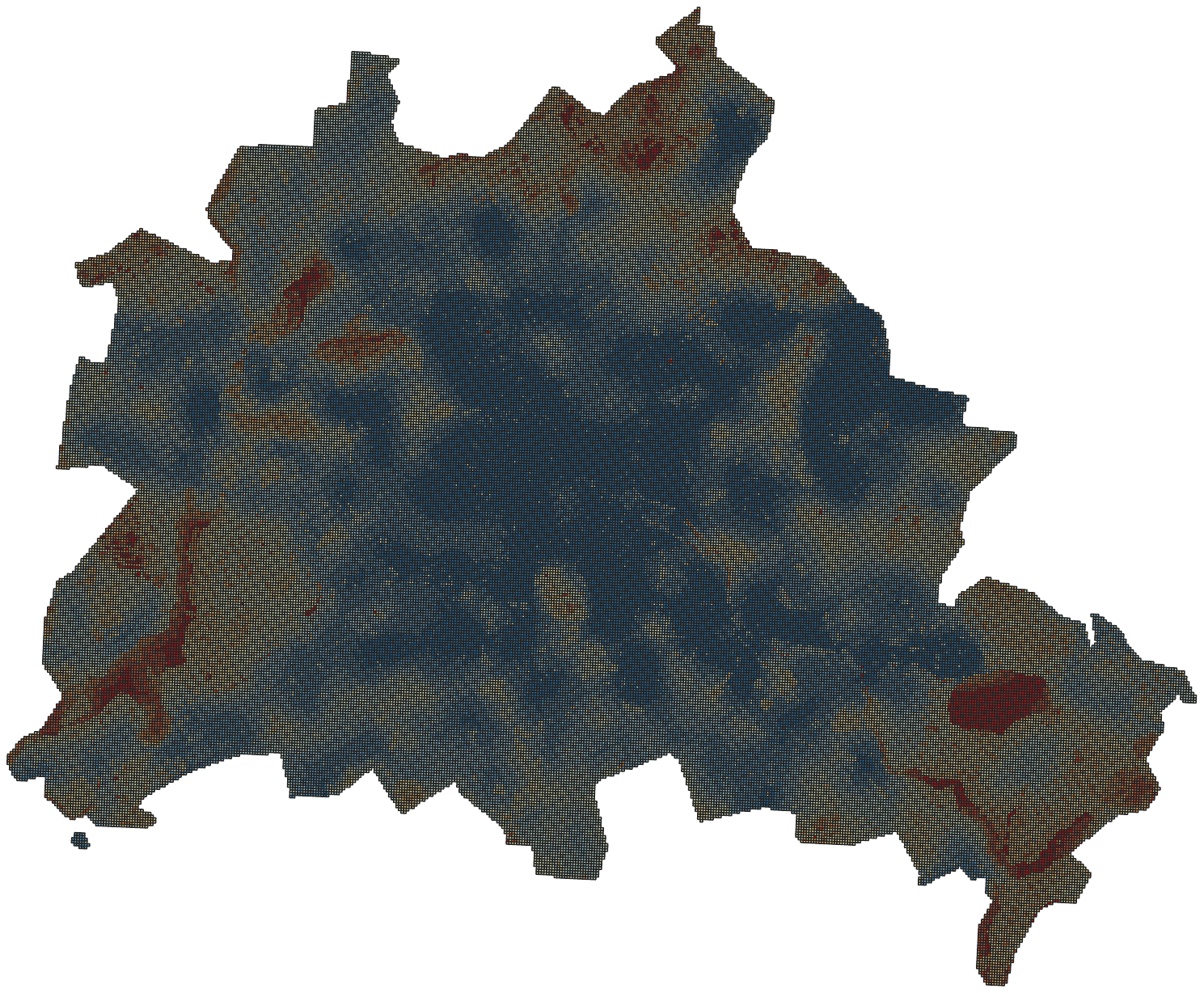}
        \caption{Berlin, $I=0.89$}
    \end{subfigure}
    \begin{subfigure}[b]{0.24\textwidth}
        \includegraphics[width=\textwidth]{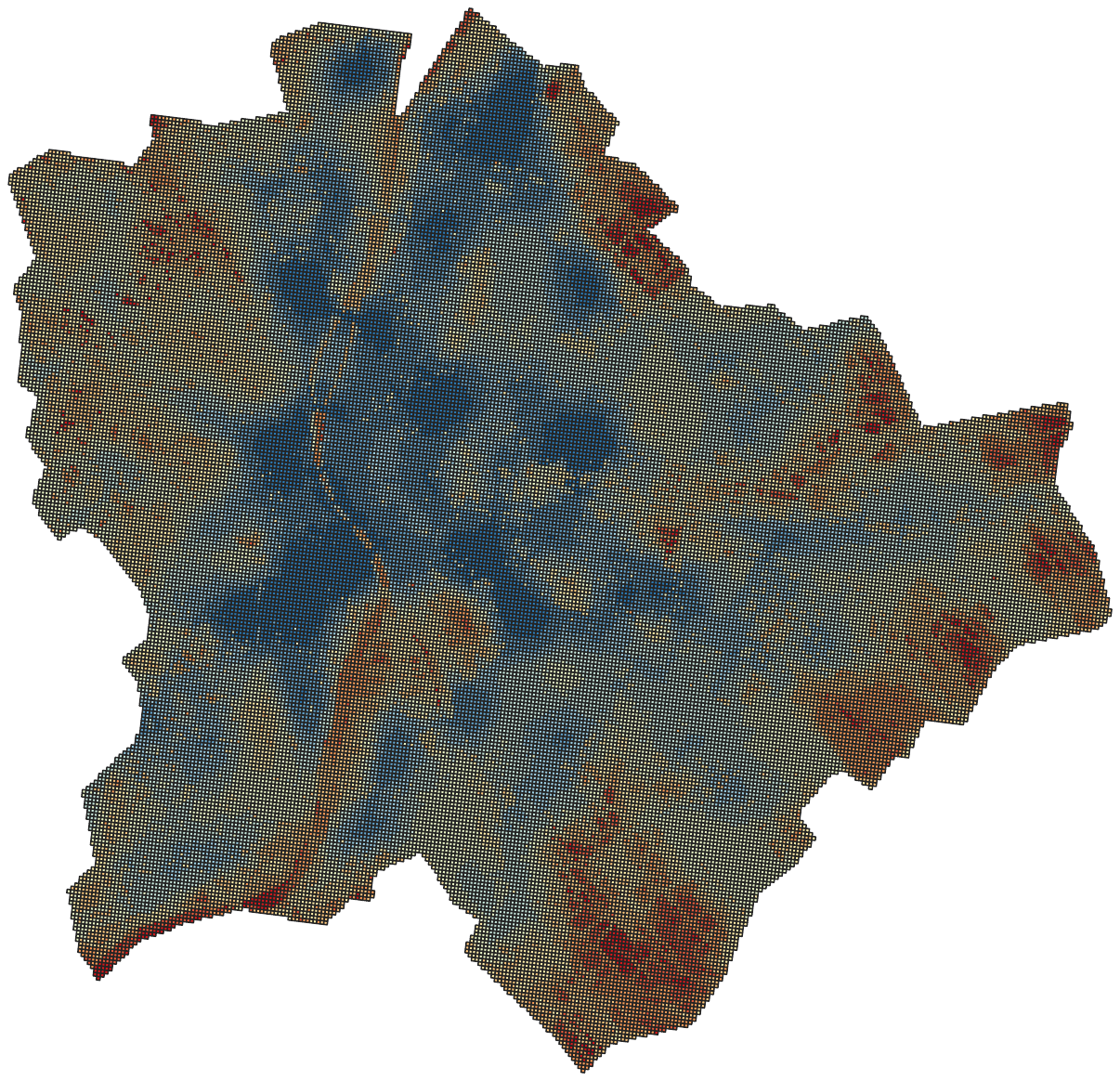}
        \caption{Budapest, $I=0.89$}
    \end{subfigure}
    
    \begin{subfigure}[b]{0.24\textwidth}
        \includegraphics[width=\textwidth]{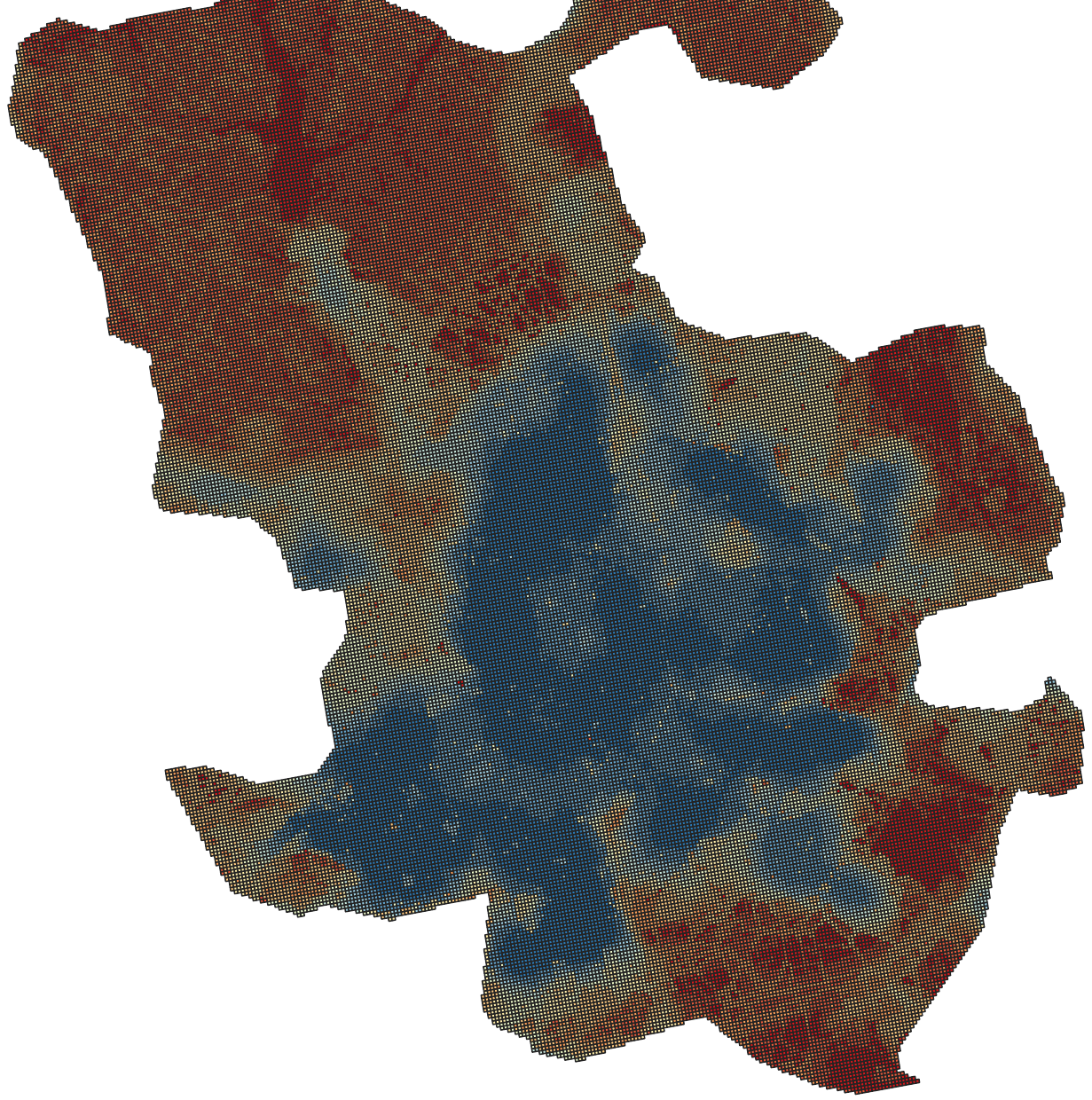}
        \caption{Madrid, $I=0.97$}
    \end{subfigure}
    \begin{subfigure}[b]{0.24\textwidth}
        \includegraphics[width=\textwidth]{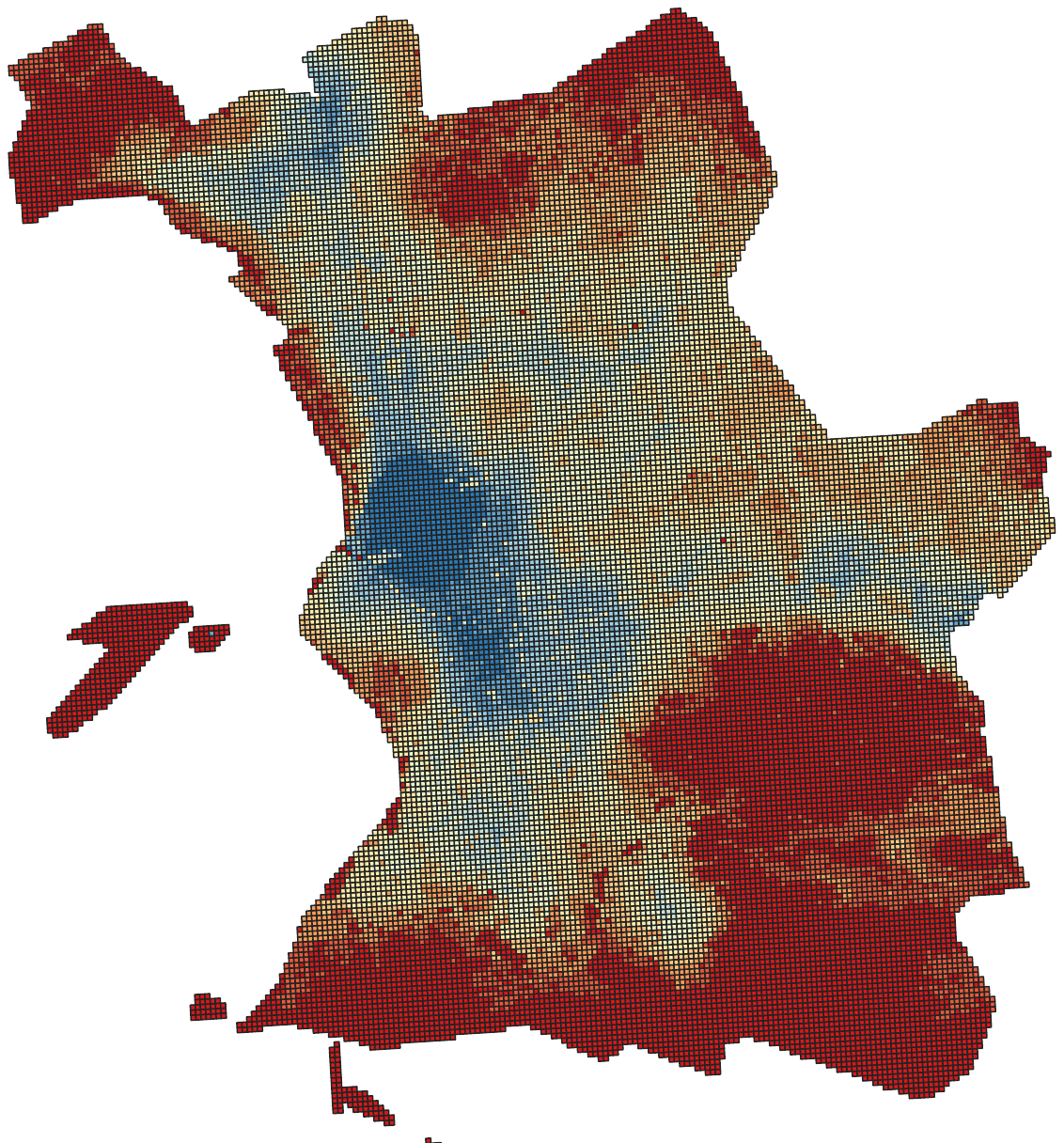}
        \caption{Marseille, $I=0.94$}
    \end{subfigure}
    \begin{subfigure}[b]{0.24\textwidth}
        \includegraphics[width=\textwidth]{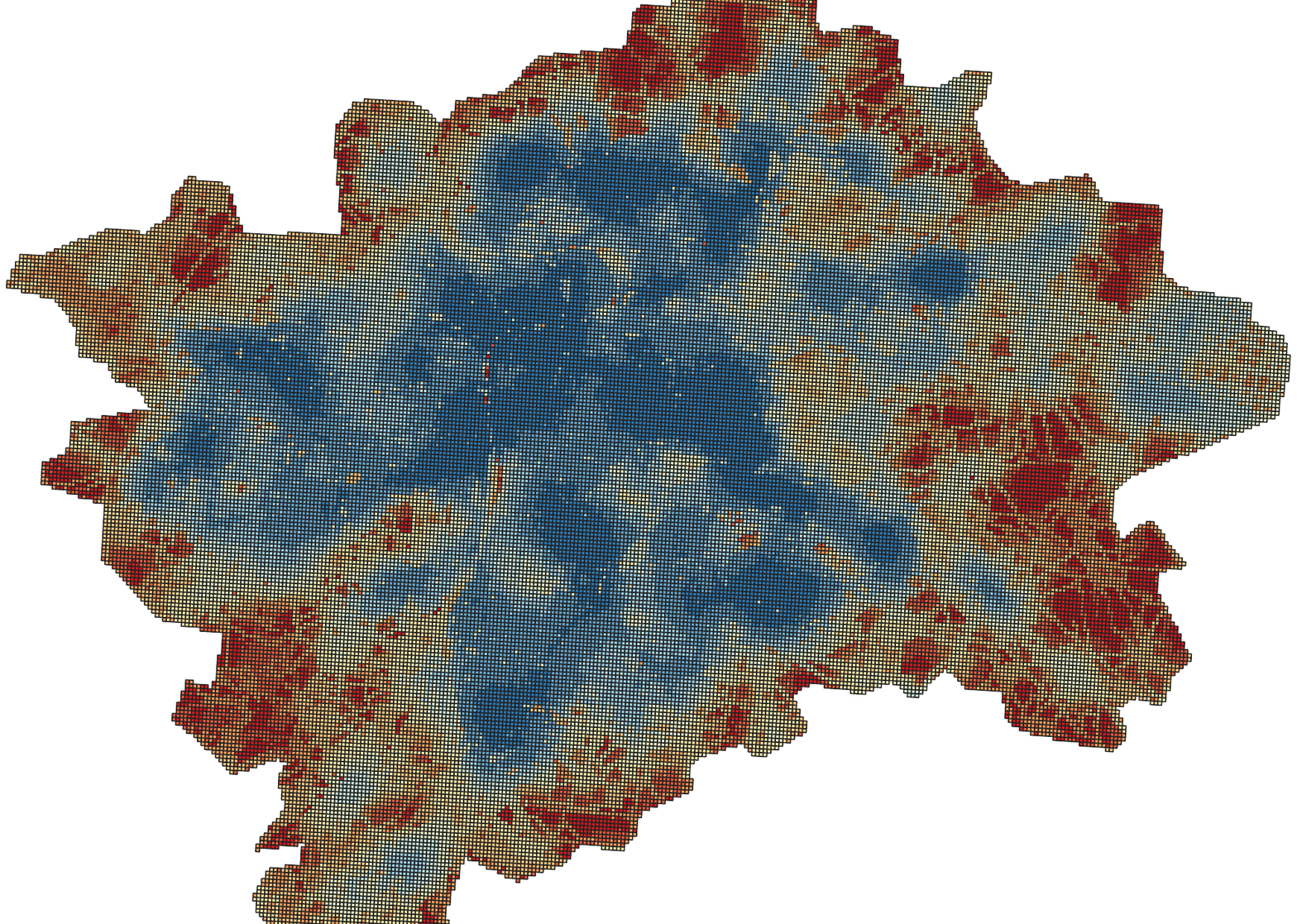}
        \caption{Munich, $I=0.89$}
    \end{subfigure}
    \begin{subfigure}[b]{0.24\textwidth}
        \includegraphics[width=\textwidth]{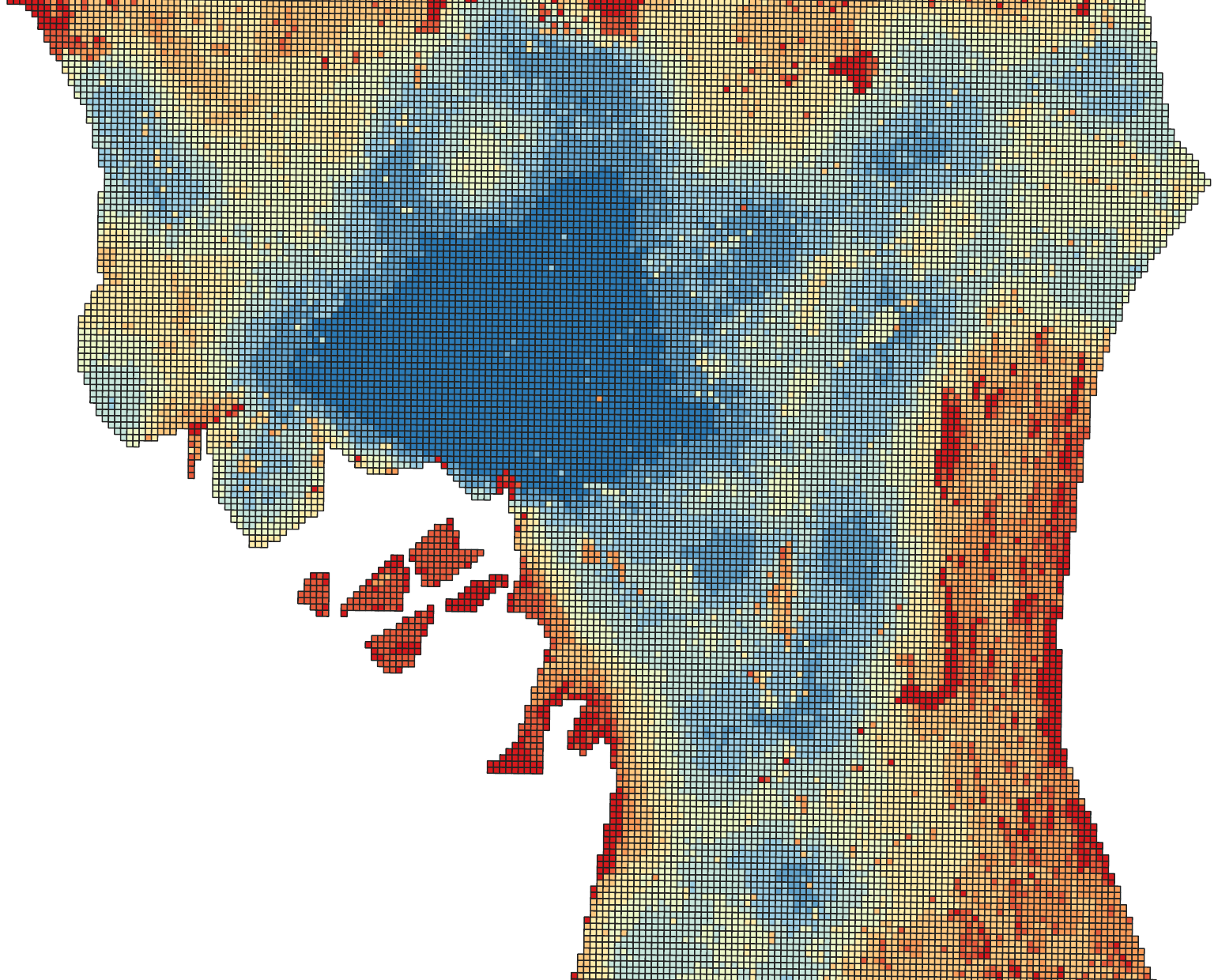}
        \caption{Oslo, $I=0.93$}
    \end{subfigure}
    
    \begin{subfigure}[b]{0.24\textwidth}
        \includegraphics[width=\textwidth]{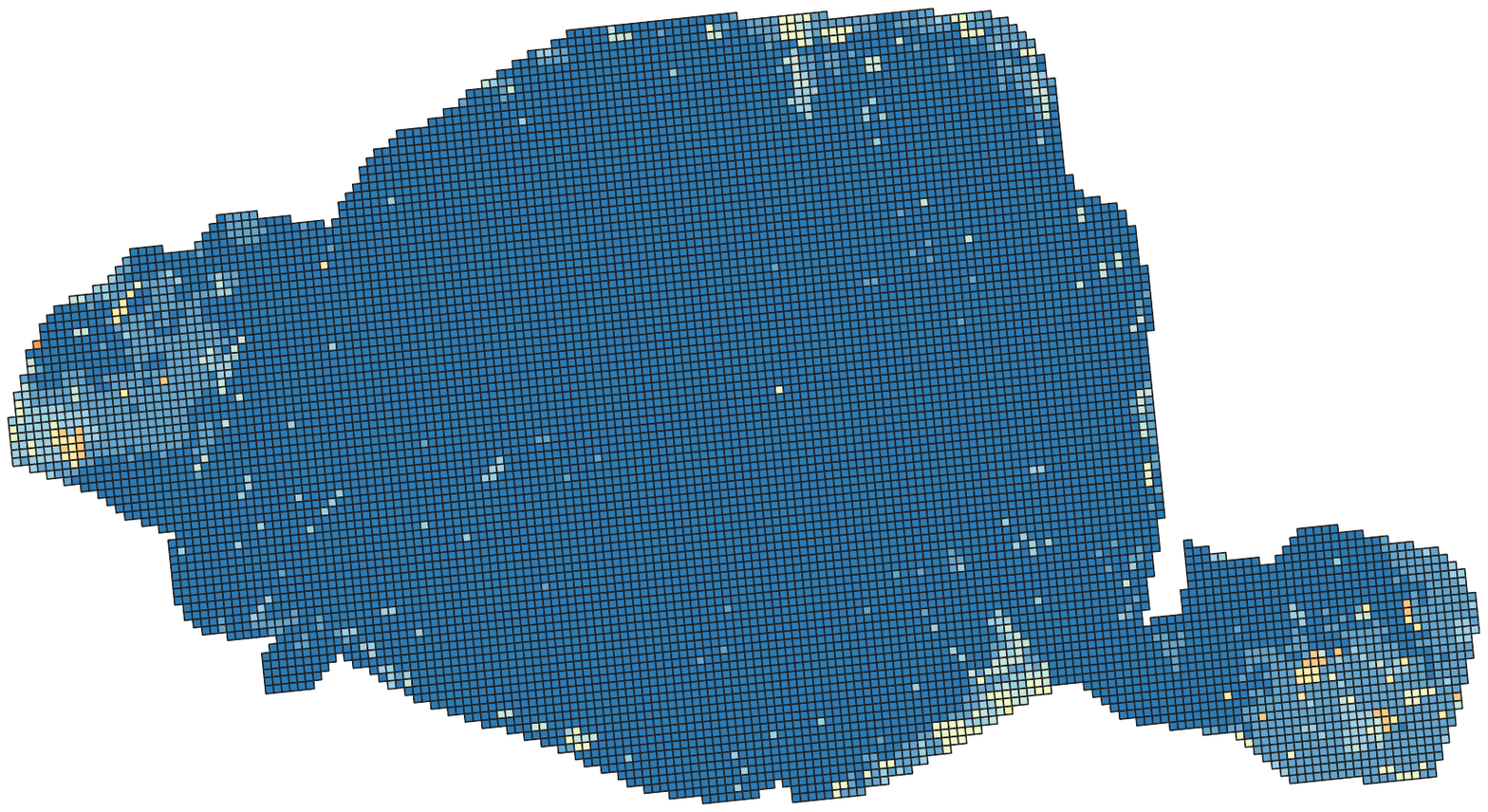}
        \caption{Paris, $I=0.87$}
    \end{subfigure}
    \begin{subfigure}[b]{0.24\textwidth}
        \includegraphics[width=\textwidth]{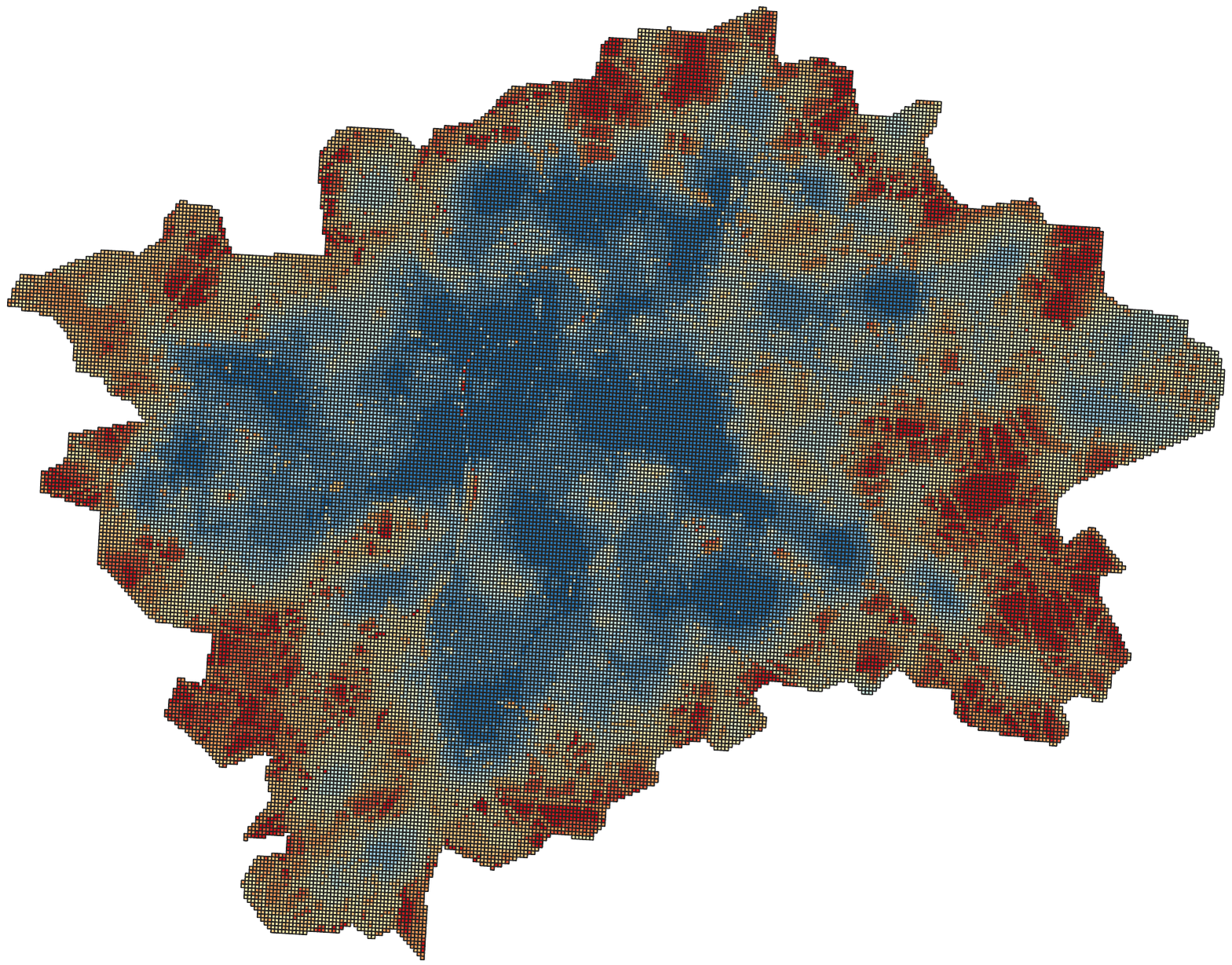}
        \caption{Prague, $I=0.91$}
    \end{subfigure}
    \begin{subfigure}[b]{0.24\textwidth}
        \includegraphics[width=\textwidth]{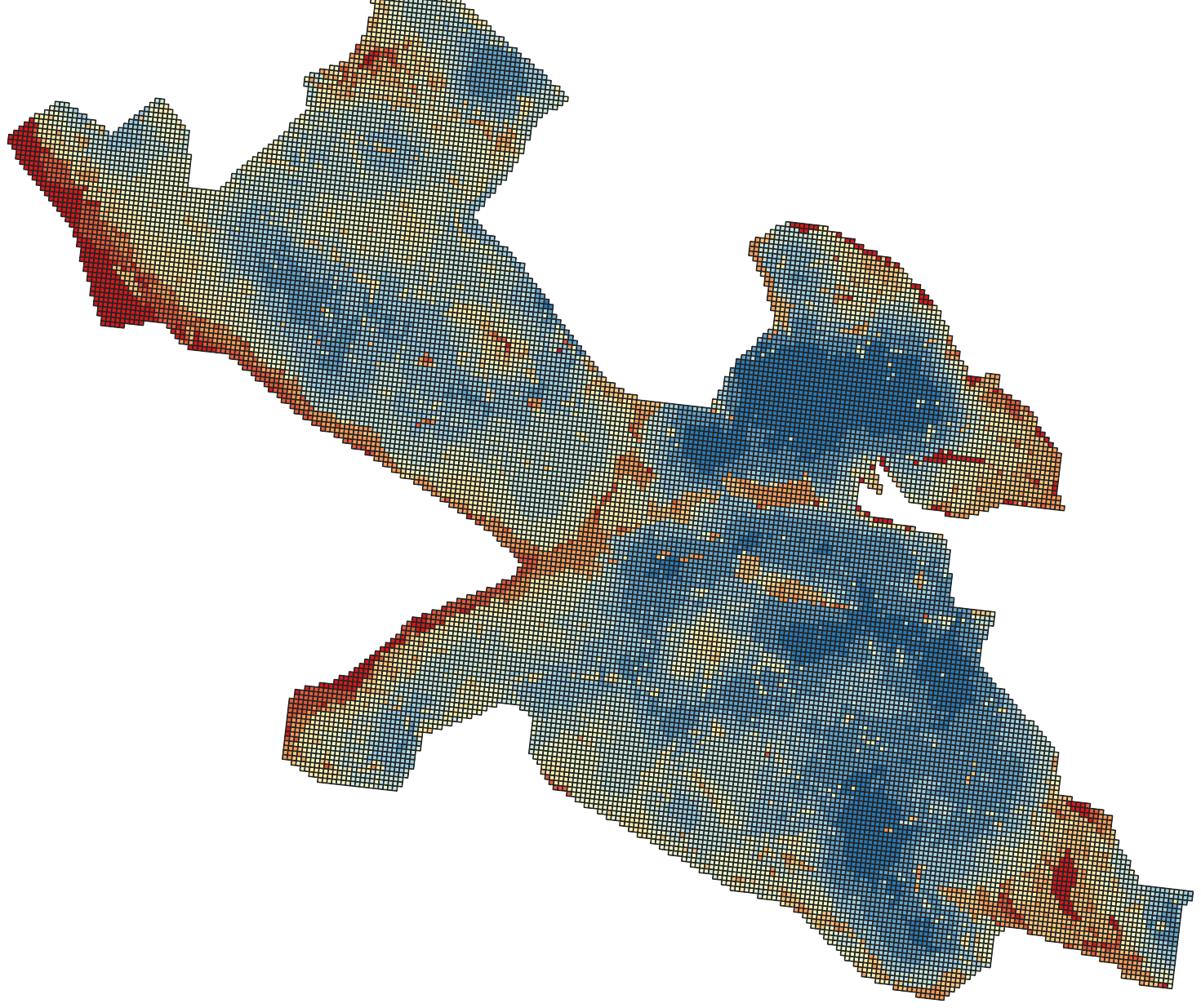}
        \caption{Stockholm, $I=0.86$}
    \end{subfigure}
    \begin{subfigure}[b]{0.24\textwidth}
        \includegraphics[width=\textwidth]{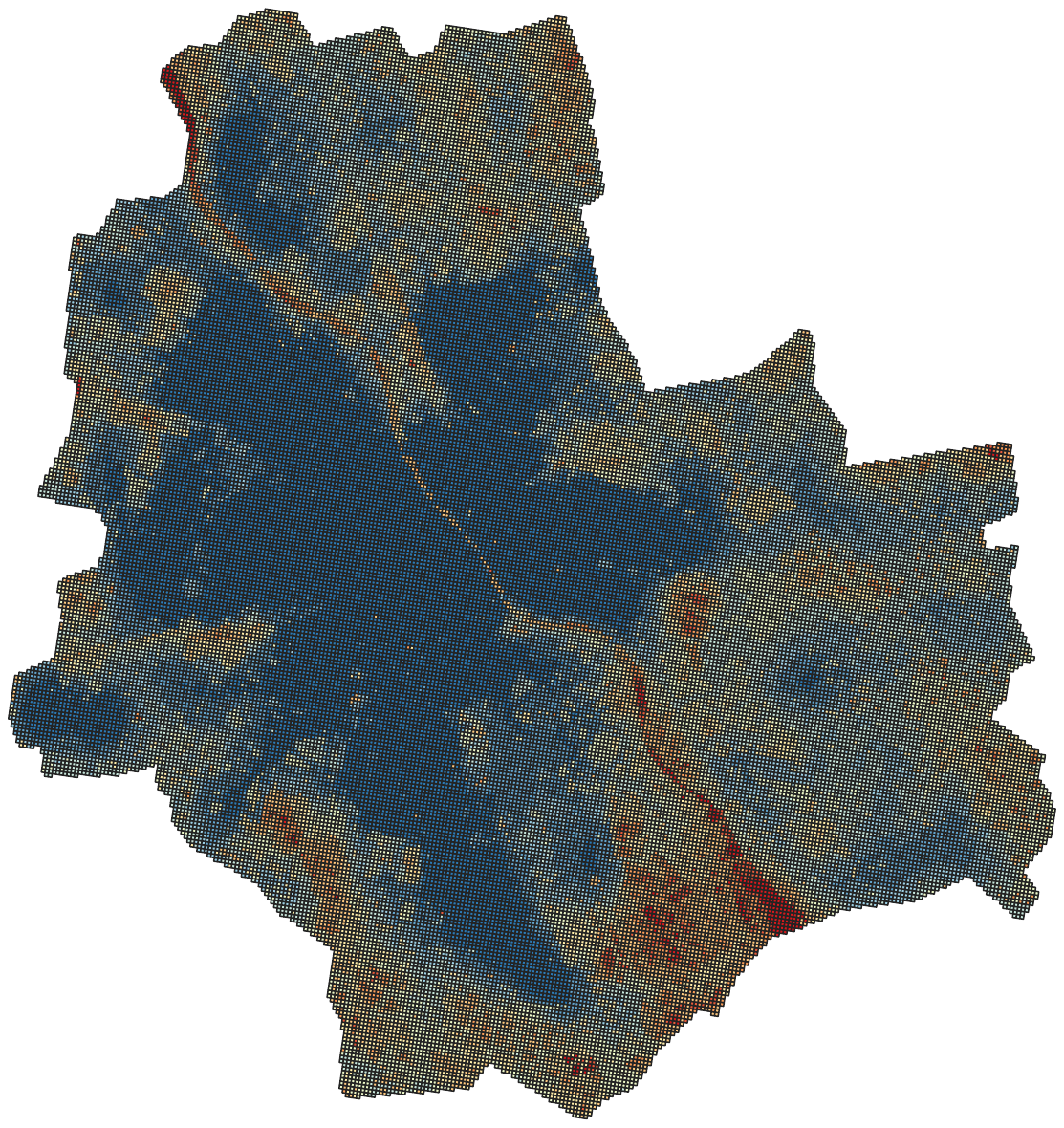}
        \caption{Warsaw, $I=0.91$}
    \end{subfigure}

    \begin{subfigure}[b]{\textwidth}
        \centering
        \includegraphics[width=0.7\textwidth]{figures_new/legend_cities.png}
    \end{subfigure}
    
    \caption{Walkability index maps of selected metropolitan cities at the original 100 m × 100 m resolution, shown alongside their corresponding global Moran's I values. \textbf{Please note}: the images are embedded as PDFs, which may appear darker or less detailed at lower zoom levels. For optimal clarity, readers are encouraged to zoom in. For high-resolution visuals and interactive exploration, visit the \href{https://ohheynish.github.io/walkability_web_atlas}{interactive web atlas}.}
    \label{fig:cities_map}
\end{figure*}

Furthermore, Figure \ref{fig:barh} displays the percent of population living below $6^\text{th}$ decile of walkability across various metropolitan cities across Europe. Figure \ref{fig:curve} illustrates the cumulative distribution of the developed walkability index across population in these metropolitan cities. Figure \ref{fig:cities_map} shows the walkability patterns of some of these metropolitan cities at the original scale of 100 m x 100 m alongside their Moran's I values calculated using Equation \eqref{eq:5}. As indicated by high values of Moran's I, there is a high spatial autocorrelation in the developed walkability index. This is expected as the development of the walkability index takes into account the values of neighboring cells, as highlighted by Equation \eqref{eq:1}.

\begin{figure*}[!t]
    \centering
    \begin{subfigure}[b]{0.32\textwidth}
        \includegraphics[width=\textwidth]{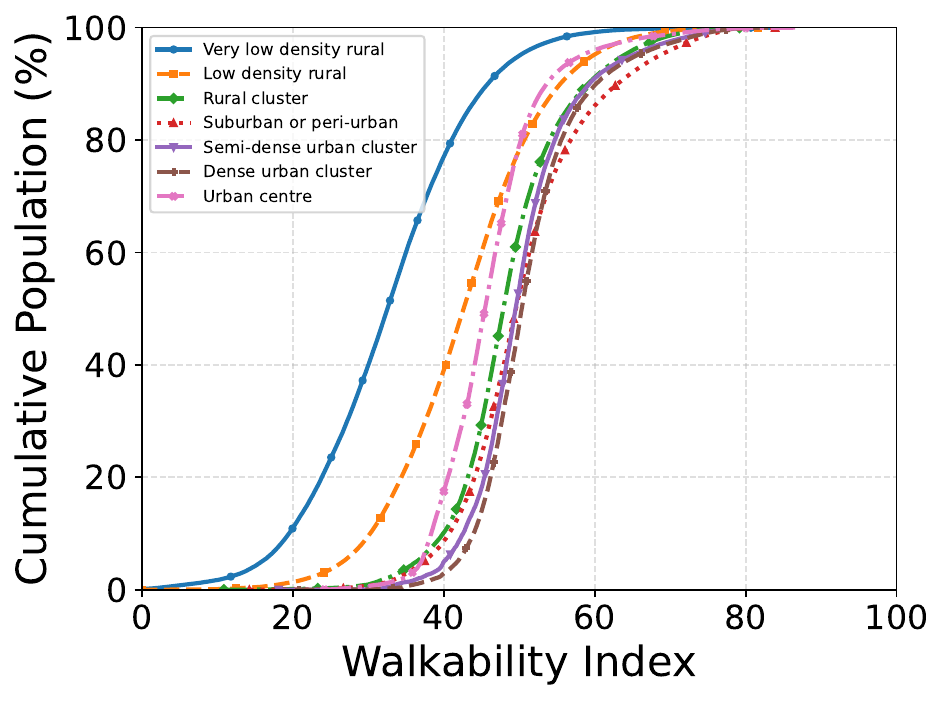}
        \caption{Austria}
    \end{subfigure}
    \begin{subfigure}[b]{0.32\textwidth}
        \includegraphics[width=\textwidth]{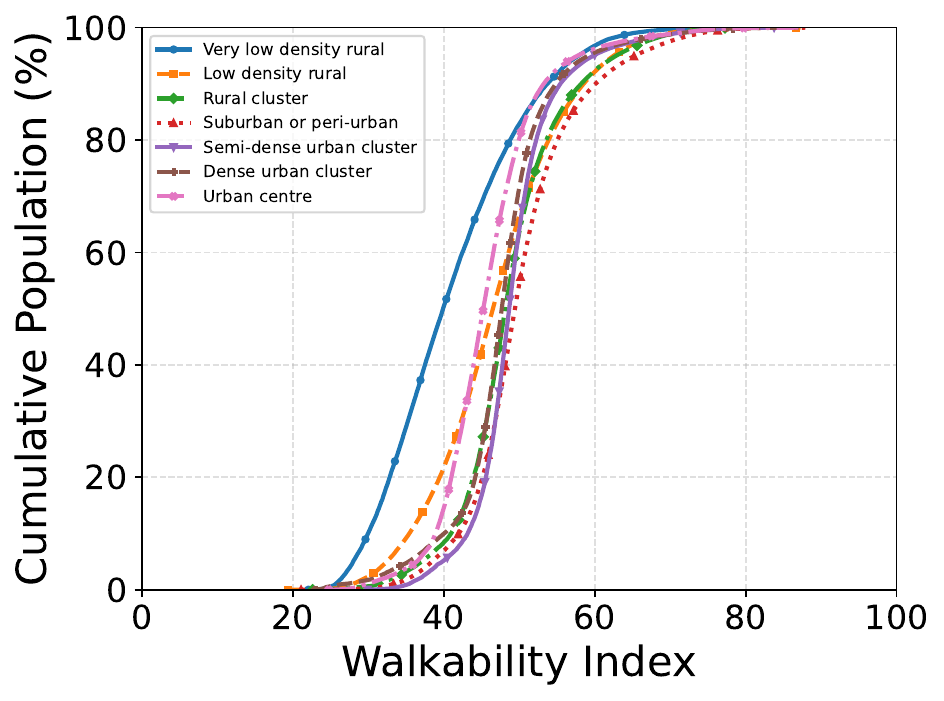}
        \caption{Belgium}
    \end{subfigure}
    \begin{subfigure}[b]{0.32\textwidth}
        \includegraphics[width=\textwidth]{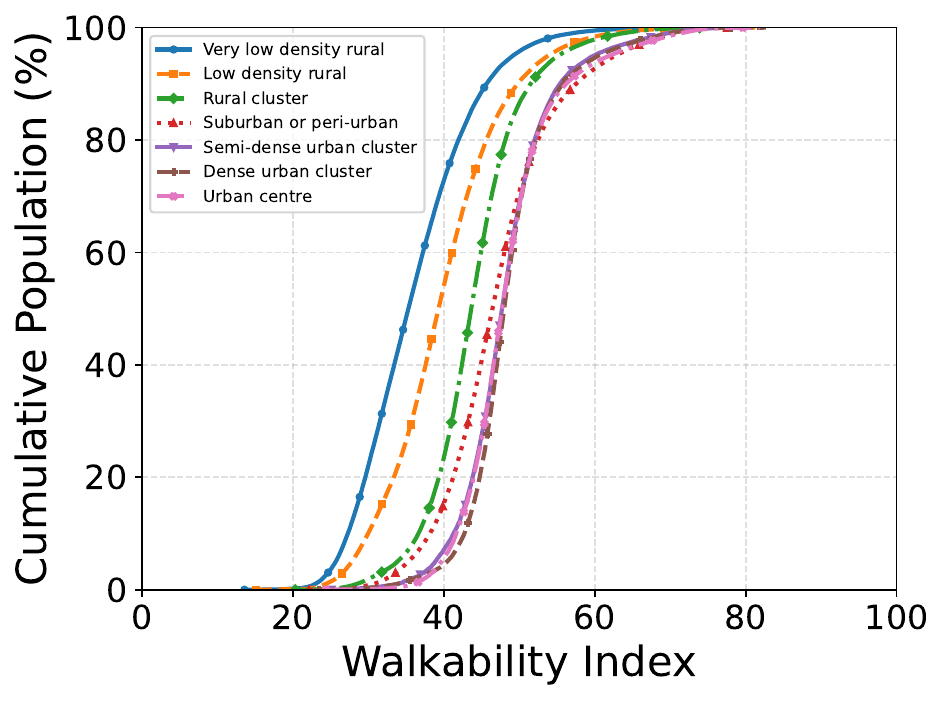}
        \caption{Czechia}
    \end{subfigure}
    
    \begin{subfigure}[b]{0.32\textwidth}
        \includegraphics[width=\textwidth]{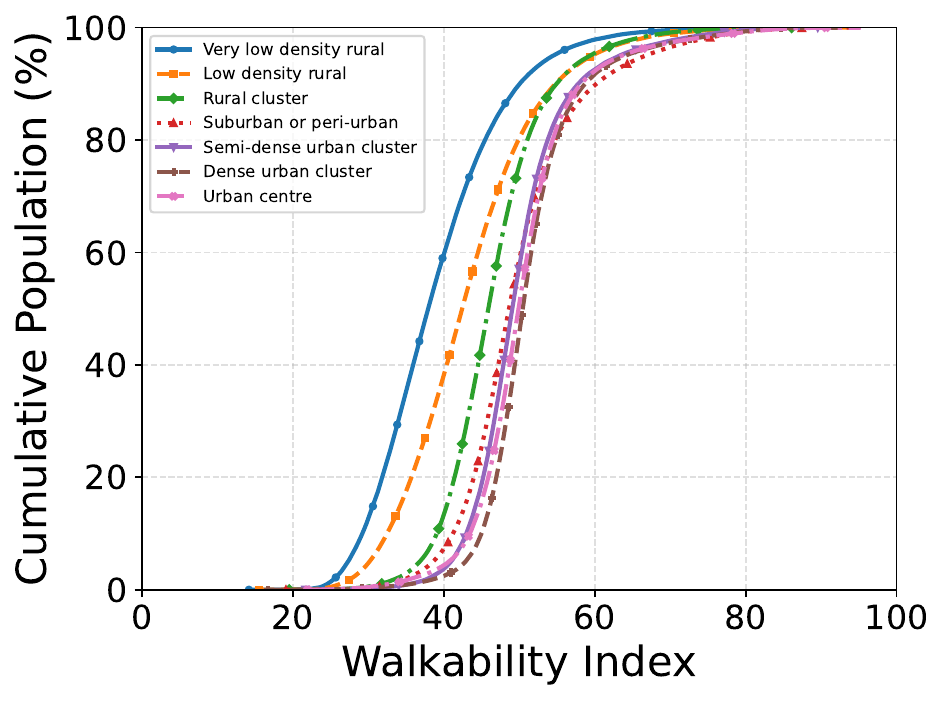}
        \caption{Germany}
    \end{subfigure}
    \begin{subfigure}[b]{0.32\textwidth}
        \includegraphics[width=\textwidth]{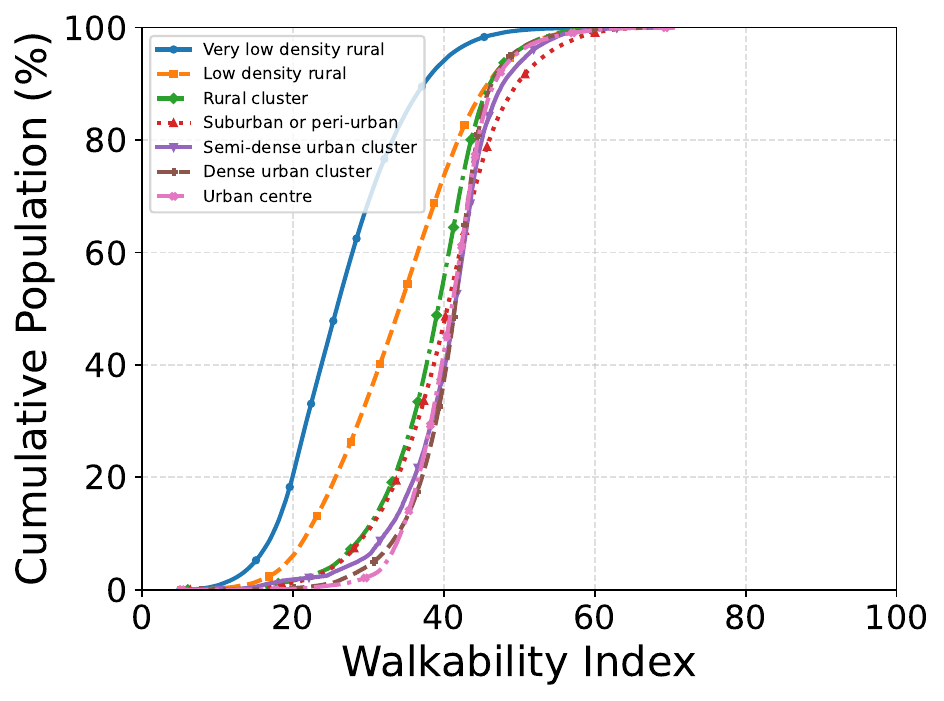}
        \caption{Spain}
    \end{subfigure}
    \begin{subfigure}[b]{0.32\textwidth}
        \includegraphics[width=\textwidth]{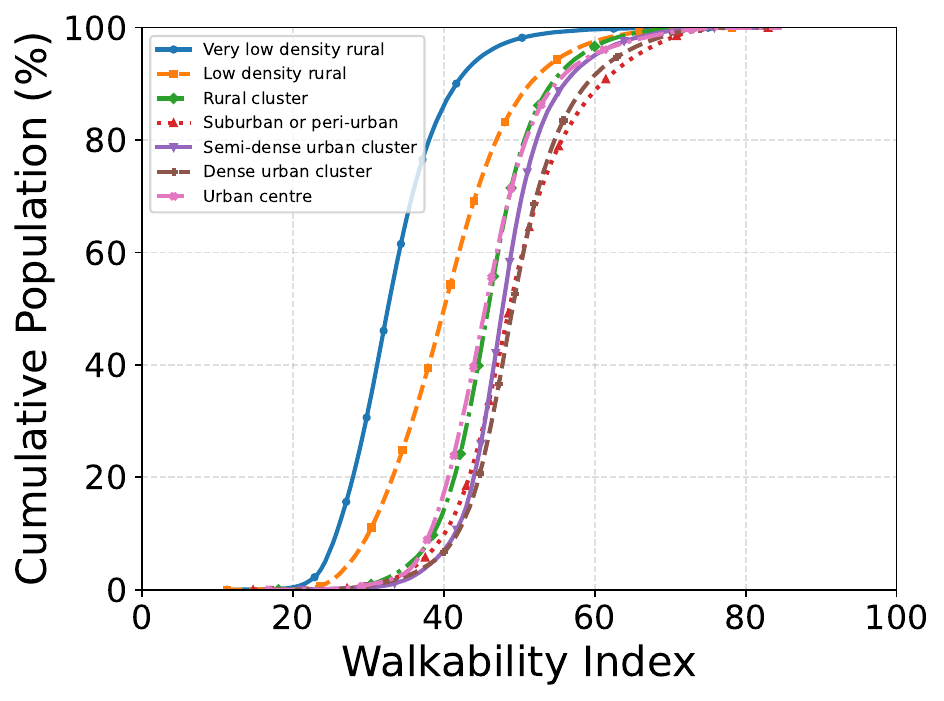}
        \caption{France}
    \end{subfigure}
    
    \begin{subfigure}[b]{0.32\textwidth}
        \includegraphics[width=\textwidth]{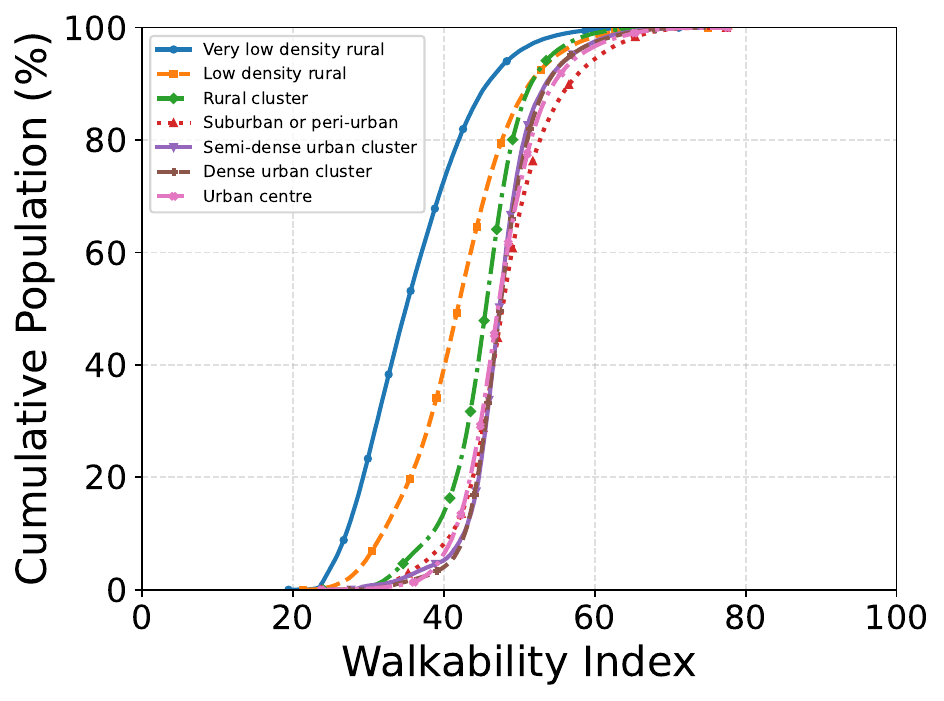}
        \caption{Hungary}
    \end{subfigure}
    \begin{subfigure}[b]{0.32\textwidth}
        \includegraphics[width=\textwidth]{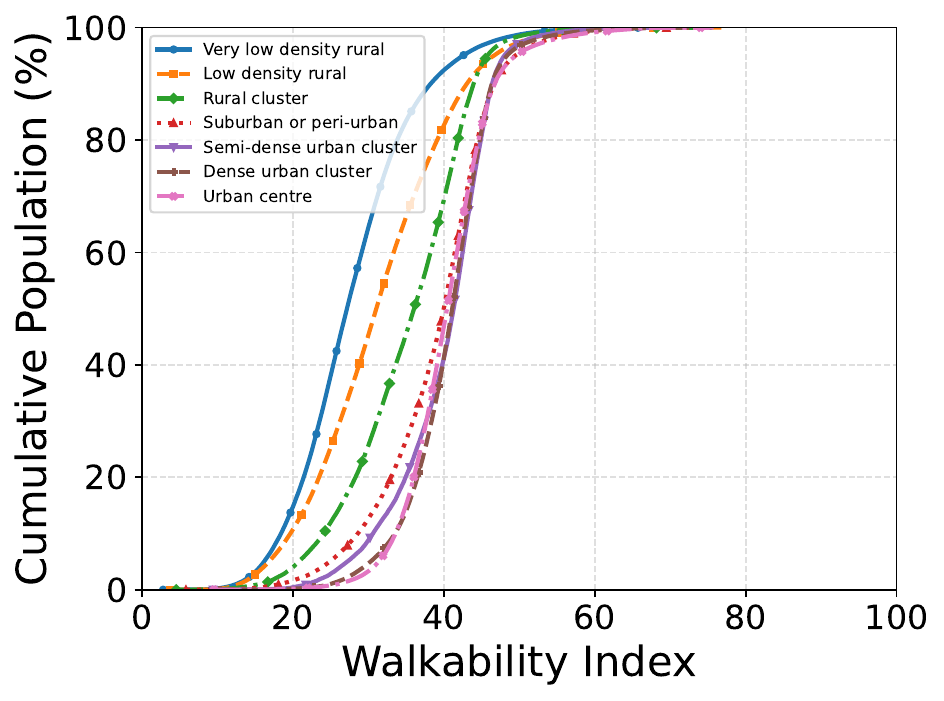}
        \caption{Italy}
    \end{subfigure}
    \begin{subfigure}[b]{0.32\textwidth}
        \includegraphics[width=\textwidth]{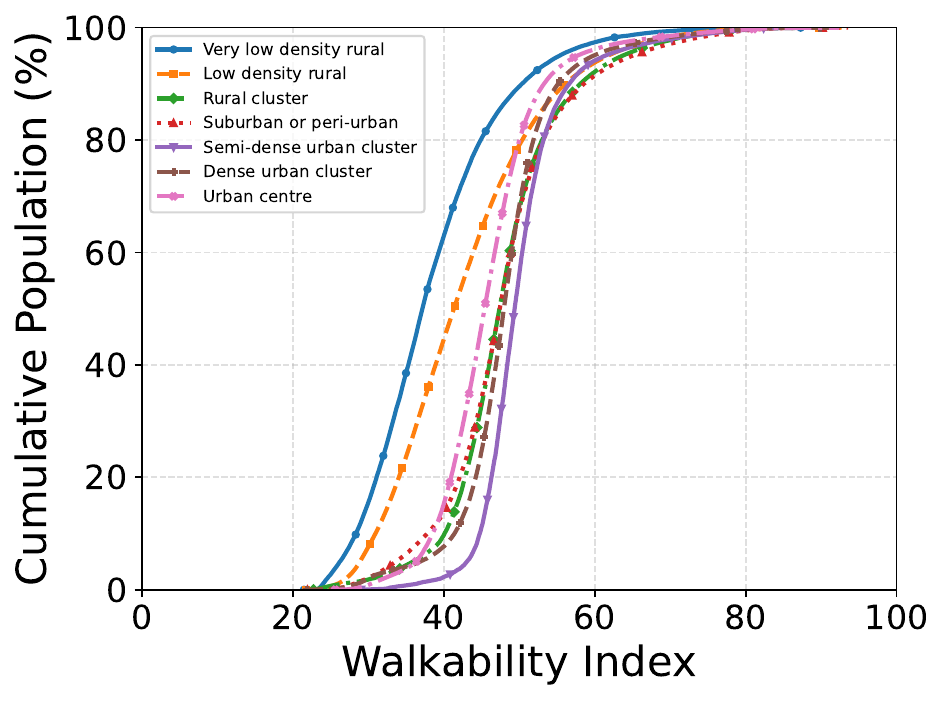}
        \caption{Netherlands}
    \end{subfigure}

    \begin{subfigure}[b]{0.32\textwidth}
        \includegraphics[width=\textwidth]{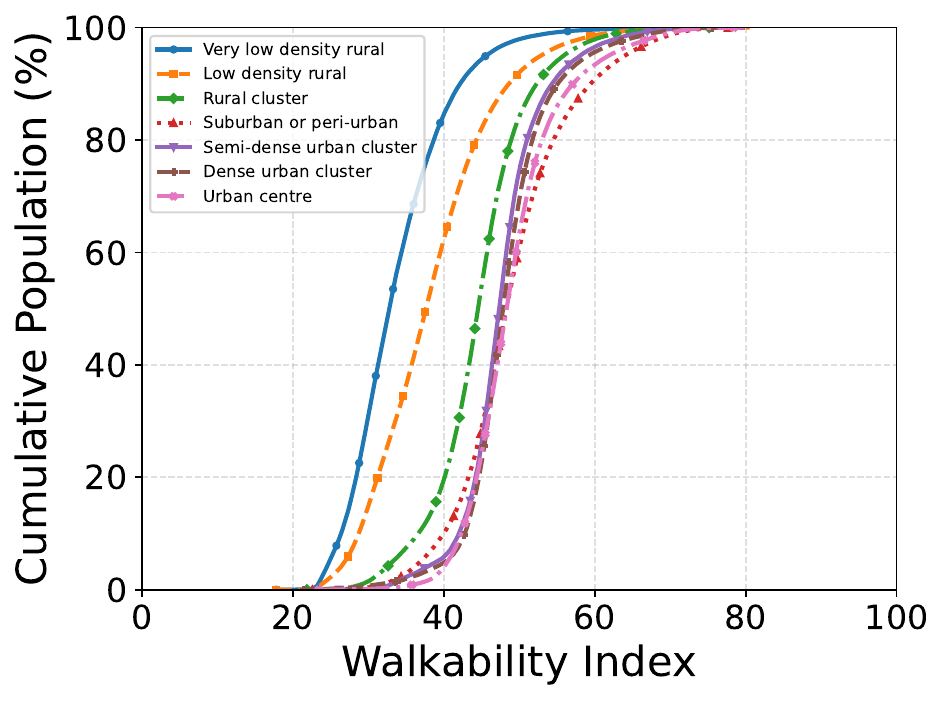}
        \caption{Poland}
    \end{subfigure}
    \begin{subfigure}[b]{0.32\textwidth}
        \includegraphics[width=\textwidth]{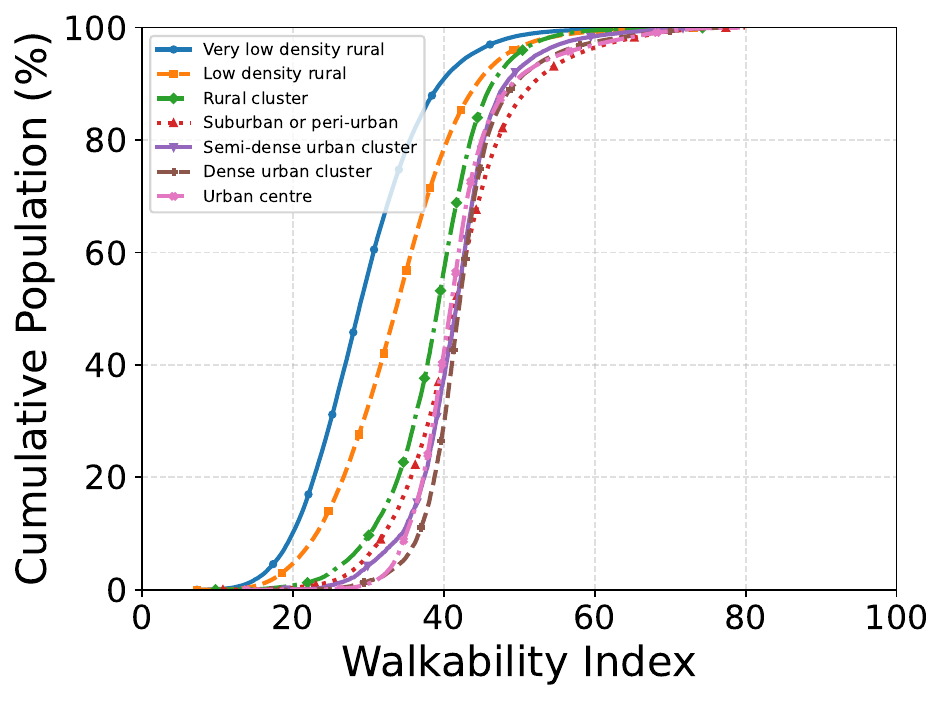}
        \caption{Portugal}
    \end{subfigure}
    \begin{subfigure}[b]{0.32\textwidth}
        \includegraphics[width=\textwidth]{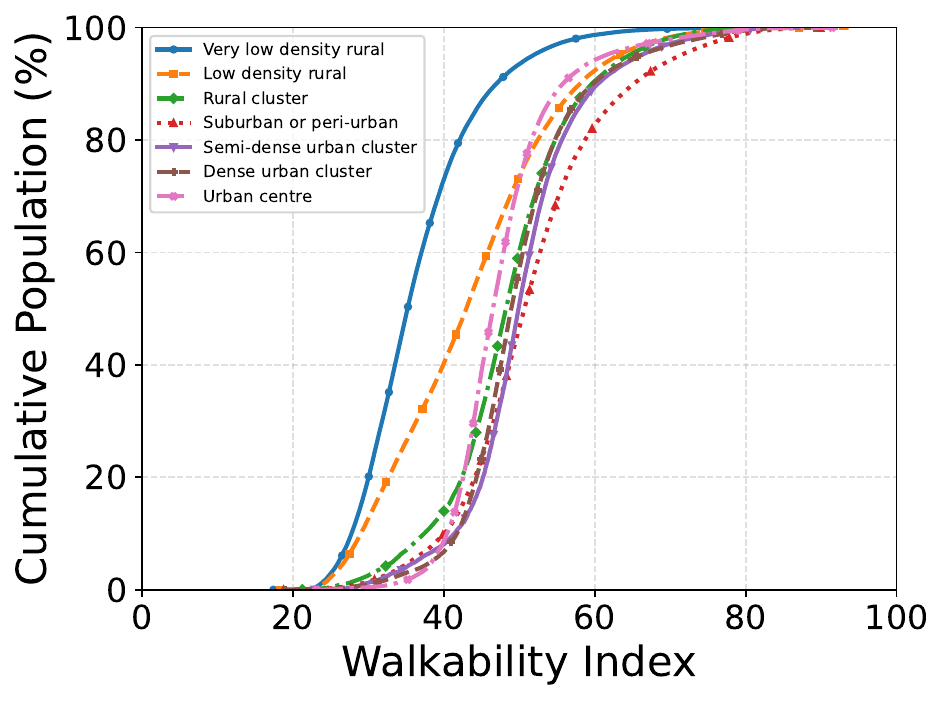}
        \caption{Sweden}
    \end{subfigure}

    \caption{Cumulative distribution of walkability index by degree of urbanization within each selected country. The curves represent the cumulative population share as a function of the walkability index, stratified by degree of urbanization. Different line styles and markers distinguish urbanization categories, ranging from very low density rural to urban centres. Countries with more steep, right-shifted curves indicate greater walkability, while flatter curves reveal lower and more varied walkability levels.}
    \label{fig:country_cdf}
\end{figure*}

Finally, Figure \ref{fig:country_cdf} shows the cumulative distribution of walkability stratified by degree of urbanization within each selected country. The curves illustrate how walkability is distributed across settlement types, from very low-density rural areas to urban centres. Countries with steeper, right-shifted curves in higher-density urban classes indicate greater access to walkable environments for a larger share of the population. In contrast, flatter or left-skewed curves—especially in rural categories—reveal greater inequality and lower overall walkability. An interesting observation emerges for countries such as Austria, Belgium, the Netherlands, and Sweden, where the cumulative distribution curves for certain rural areas are slightly more right-shifted than those of urban centres. This pattern deviates from the expected urbanization gradient seen in Figure~\ref{fig:strip_plot}, where walkability typically increases with urban density. One possible explanation lies in the per capita adjustment applied to several components of the index in Figure~\ref{fig:country_cdf}.

\section{Discussion}
\subsection{Strengths}\label{strengths}
One of the key strengths of this study is the spatial coverage, while simultaneously increasing the level of detail. The high granularity of mapping allows for the identification of subtle, localized environmental differences that can have a meaningful impact on physical activity behaviors and further health outcomes. Furthermore, the standardization of methods and input variables across large areas enables comparisons across local areas, countries, and European regions, offering valuable insights for sub-national, national, regional, and EU-wide policy initiatives aimed at improving public health through urban planning. By providing open access to our code, we enable researchers to generate alternative versions of the index for different geographic regions, modify key parameters to model various aspects of urban environments, and test different hypotheses regarding walkability and its implications in a relatively standardized manner. This open access approach improves the diagnostic utility of the index, particularly when used as an exposure variable in studies linking walking to health outcomes such as obesity and physical activity levels. Furthermore, the interactive map we provide can serve as an essential resource for urban planners, policymakers, and researchers interested in exploring spatial patterns of walkability at both macro and micro scales. By facilitating comparative analysis across diverse geographic contexts at scale, this study contributes to a broader understanding of how the built environment influences mobility and active lifestyles.  

\subsection{Limitations}\label{limitations}
While this study implements nuanced techniques such as street network buffers and distance-decayed kernels at scale to construct a granular walkability measure, it still exhibits both theoretical and practical constraints that must be acknowledged. Theoretically, one key limitation is the lack of rigorous justification for the environmental components associated with walkability, particularly at scale. While local-scale studies often rely on region-specific knowledge or surveys to justify these components, such an approach is impractical for a large-scale study like this. As a result, there is an inherent trade-off between generalization and local relevance — standardized measures enable broad comparisons but may overlook critical local factors, such as cultural or climatic influences on walking behavior. 

Another theoretical challenge is the reliance on OSM as a data source. While OSM provides a standardized and scalable dataset, its completeness and quality varies by region, particularly in less-documented areas where contributions are sparse \cite{OSM_Completeness}. This could potentially introduce bias and uncertainty in the data and can affect the reliability of the components of the walkability index derived from OSM data (\texttt{SWL}, \texttt{SI}, \texttt{PT}, and \texttt{ISO}). While prior validation studies have shown that OSM data achieves relatively high completeness in many European urban areas \cite{barrington2017world}, the potential for spatial bias remains, particularly in rural or less developed regions. Despite these limitations, OSM presents several important advantages for a study of this geographic and thematic scale. It is among the few openly available, globally consistent sources of spatial data. In the absence of harmonized official datasets across countries and regions, OSM remains a valuable resource for conducting comparative walkability assessments. Furthermore, as OSM continues to evolve and become more complete, future iterations of the index can benefit from improved data quality. Our fully open-source workflow ensures that the index is both reproducible and easily updatable as OSM data improves over time.

Additionally, certain urban components, such as \texttt{SWL}, \texttt{SI}, and \texttt{ISO}, exhibit high correlation, which may distort the index if their weights are not properly adjusted. However, existing literature lacks a robust methodology for determining these weights, posing another challenge \cite{venerandi2024walkability}. One potential approach is to assign weights inversely proportional to each component's total correlation with others. This data-driven method downweights redundant components and could offer a useful alternative to equal or expert-based weighting.

From a practical standpoint, this study faces significant computational challenges in processing data for $\sim 1$ billion neighborhoods. Generating NDVI images from Sentinel-2 data alone require approximately two weeks of processing time on Google Earth Engine, which varies unpredictably based on server traffic. Similarly, collecting and processing OSM data for all of Europe takes several days, depending on the component. While GPU-based algorithms accelerate distance-decay computations, data preparation for GPU processing remains a bottleneck. Constructing street network buffers using the Valhalla routing engine also introduces inefficiencies due to network instability, occasionally leading to failed isochrone requests.

\subsection{Future Works}
\subsubsection{Validation and Ground Truth} \label{validation}
Future research should focus on several key areas to further enhance the utility and applicability of the current walkability index. First, validation with walking and also physical activity data, especially objectively measured using personal sensors, is essential to establish stronger empirical links between walkability and actual walking habits and then physical activity. Second, examining the links between walkability and downstream health outcomes including obesity, chronic diseases, and mental health, would provide insights into its broader public health implications. Third, integrating this walkability index into broader composite indices that capture a wider range of environmental aspects related to health behaviors beyond physical activity could offer a more comprehensive understanding of environmental influences on health. For e.g., combining it with constructs capturing the healthiness of the food environment, drivability, bikeability and sports facilities can together capture the ‘obesogenicity’ or healthfulness of environments as has been done in more local settings \cite{Cebrecos2019-te, Kaczynski2020-cn, Marek2021-pi, lam2022development}. Additionally, adapting methods to capture localized context, such as cultural, socio-economic, and climatic factors, could improve the relevance and accuracy of the index in diverse settings. Another important area for future work is the validation of OSM derived components used in the index. Although OSM provides a rich and scalable dataset for capturing street networks, land use, and points of interests (POIs), its completeness can vary substantially across regions. Comparing OSM based features with official governmental mapping sources, such as National Mapping Agencies or local geographic datasets, could help quantify and adjust for regional data gaps or inconsistencies. Such cross validation would be especially valuable in areas with lower OSM contributor activity, ensuring a more robust and accurate representation of walkability inputs across geographies.

\subsubsection{Use of foundational AI models}
As discussed throughout this paper, data sources such as OSM, satellite imagery, and other secondary geospatial datasets contribute to the standardization of definitions and large-scale processing of environmental components while maintaining granularity. However, reducing this rich environmental "big data" into a limited set of components to explore relationships with downstream health outcomes may result in the loss of important contextual signals. Recent advancements in artificial intelligence (AI) applied to Earth observation (EO) and geospatial data — including foundational models such as SatCLIP \cite{klemmer2024satclipglobalgeneralpurposelocation} and the Google Population Dynamics Model \cite{agarwal2025generalgeospatialinferencepopulation} — offer novel opportunities to bridge this gap. These models can be used to link EO-derived spatial information, as well as data from sources like Google Maps, weather records, and air-quality monitoring, to walkability and, more directly, to downstream health outcomes. Their core principle lies in generating robust environmental representations that capture both global and local variations across space directly from the input big data. To predict walkability, these models can be fine-tuned using a relatively small but spatially diverse set of ground-truth data, such as physical activity measurements or rigorously developed walkability indices that incorporate expert input and public participation. This further strengthens the argument for developing robust methods to collect ground-truth data, as discussed in \ref{validation}.
 
\section{Conclusion}
This study leveraged standardized datasets — including OSM, satellite imagery, and other secondary geospatial sources — combined with advanced techniques such as street network-based buffers and distance decay kernels to characterize walkability across Europe. The resulting maps were then utilized to analyze disparities in walkability using population data both within and between countries, regions, and cities. The high-resolution, standardized walkability index developed in this work offers an accessible resource for researchers, policymakers, and urban planners, supporting efforts to promote active living and enhance public health across diverse European settings.

\section{Funding}
This work is conducted within the OBCT project (\url{www.obct.nl}) which has received funding from the European Union's Horizon Europe research and innovation programme under grant agreement No. 101080250. Views and opinions expressed are however those of the author(s) only and do not necessarily reflect those of the European Union. Neither the European Union nor the granting authority can be held responsible for them. ISGlobal acknowledges support from the grant CEX2023-0001290-S funded by MCIN/AEI/ 10.13039/501100011033, and support from the Generalitat de Catalunya through the CERCA Program.

\bibliographystyle{apalike} 
\bibliography{reference.bib}

\end{document}